\documentclass[prc,aps,twocolumn,showpacs,showkeys,amsmath,amssymb,floatfix]{revtex4}
\usepackage{booktabs,graphicx,verbatim}
\usepackage[colorlinks=true,citecolor=blue,filecolor=blue,linkcolor=blue,urlcolor=blue,pdftex]{hyperref}
\newcommand{\1}[1]{\, \mathrm{#1}} % unit(y ;-)
\newcommand{\n}[1]{\mathrm{#1}}    % normal (roman) text in math mode

\newcommand{\columbia}{\affiliation{Physics Department, Columbia University, New York, NY 10027, USA}}

\begin{document}

\title{New Measurement of the Scintillation Efficiency of Low-Energy Nuclear Recoils in Liquid Xenon}

\author{G.~Plante}\email[]{guillaume.plante@astro.columbia.edu}
\author{E.~Aprile}
\author{R.~Budnik}
\author{B.~Choi}
\author{K.-L.~Giboni}
\author{L.~W.~Goetzke}
\author{R.~F.~Lang}
\author{K.~E.~Lim}
\author{A.~J.~Melgarejo Fernandez}

\columbia

\begin{abstract}
	Particle detectors that use liquid xenon (LXe) as detection medium are among the leading technologies in
	the search for dark matter weakly interacting massive particles (WIMPs). A key enabling element has been
	the low-energy detection threshold for recoiling nuclei produced by the interaction of WIMPs in LXe
	targets. In these detectors, the nuclear recoil energy scale is based on the LXe scintillation signal and
	thus requires knowledge of the relative scintillation efficiency of nuclear recoils,
	$\mathcal{L}_{\n{eff}}$. The uncertainty in $\mathcal{L}_{\n{eff}}$ at low energies is the largest
	systematic uncertainty in the reported results from LXe WIMP searches at low masses. In the context of the
	XENON Dark Matter project, a new LXe scintillation detector has been designed and built specifically for
	the measurement of $\mathcal{L}_{\n{eff}}$ at low energies, with an emphasis on maximizing the
	scintillation light detection efficiency to obtain the lowest possible energy threshold. We report new
	measurements of $\mathcal{L}_{\n{eff}}$ at low energies performed with this detector. Our results suggest
	a $\mathcal{L}_{\n{eff}}$ which slowly decreases with decreasing energy, from $0.144 \pm 0.009$ at 15~keV
	down to $0.088^{+0.014}_{-0.015}$ at 3~keV.
\end{abstract}



\maketitle

\section{Introduction}

% liquid xenon dark matter detectors
In recent years, liquid xenon (LXe) particle detectors
\cite{Angle:2007uj,Aprile:2010um,Alner:2007ja,Lebedenko:2009xe,Minamino:2010zz} have achieved a large increase
in target mass and a simultaneous reduction in backgrounds and are now among the leading technologies in the
search for dark matter weakly interacting massive particles (WIMPs). The XENON100 experiment, with a
target mass of 62~kg and a measured electronic recoil background of $< 10^{-2} \1{events/kg/d/keV}$
\cite{Aprile:2011vb} is currently the most sensitive dark matter search in operation. Key to the performance
of XENON100 and of future experiments based on the same principle, is the ability to detect low-energy
recoiling nuclei in LXe. Since WIMPs are expected to interact primarily with atomic nuclei,
the nuclear recoil energy scale is based on the LXe direct scintillation signal and thus requires knowledge of
the scintillation yield of nuclear recoils. The uncertainty in the nuclear recoil energy scale at low energies
is the largest systematic uncertainty in the reported results from LXe WIMP
searches\cite{Aprile:2010um,Aprile:2011hx}.

% leff
The scintillation yield, defined as the number of photons produced per unit energy, depends on the type of
particle as well as the energy~\cite{Lindhard:1961zz,Hitachi:2005ti}.
Since a precise measurement of the absolute scintillation yield is rather difficult,
the relative scintillation efficiency of nuclear recoils,
$\mathcal{L}_{\n{eff}}$, is the quantity that is used to convert the scintillation signals of LXe dark matter
detectors in nuclear recoil energies. $\mathcal{L}_{\n{eff}}$ is defined as the ratio of the scintillation
yield of nuclear recoils to that of electronic recoils from photoabsorbed 122~keV $\gamma$ rays from a
$^{57}\n{Co}$ source, at zero electric field. The scintillation yield of electronic recoils is also an
energy-dependent quantity~\cite{Barabanov:1987fj}, but the 122~keV $\gamma$ rays provide a reference with which
scintillation yields at other energies or of other particles can be compared.

% theoretical attempts
%\todo{Theoretical attempts to describe the low-energy behavior of $\mathcal{L}_{\n{eff}}$ are limited, due to the
%poor knowledge of the energy loss sharing between nuclear and electronic collisions. Only part of the
%electronic excitation energy can be used for ionization and/or scintillation. The fraction of the total
%energy given to electronic excitation is the nuclear quenching factor or the Lindhard factor\ldots}

Two methods have been used to measure $\mathcal{L}_{\text{eff}}$ at different energies, an indirect method
with the full spectrum comparison of the simulated response to the measured response from the irradiation with
a neutron source, and a direct method with monoenergetic neutron fixed-angle scatters.

% indirect measurements
Indirect measurements~\cite{Sorensen:2008ec,Lebedenko:2008gb} infer the energy dependence of
$\mathcal{L}_{\text{eff}}$ by comparing experimental data obtained with a neutron source and a Monte Carlo
simulation of the expected nuclear recoil energy spectrum. In this way,
any neglected factors will be absorbed in the energy dependence of $\mathcal{L}_{\n{eff}}$. Such
factors can include uncertainties in the energy spectrum of the neutron source, efficiency losses near
threshold, energy dependence of selection cuts, etc, and are typically difficult to measure precisely.

% direct measurements
Direct measurements
\cite{Arneodo:2000vc,Bernabei:2000aa,Akimov:2001pb,Aprile:2005mt,Chepel:2006yv,Aprile:2008rc,Manzur:2009hp}
are performed by recording fixed-angle elastic scatters of monoenergetic neutrons tagged by organic liquid
scintillator detectors. The recoil energy of the Xe nucleus is then entirely fixed by kinematics and
approximately given by
\begin{equation}
	E_r \approx 2 E_n \frac{m_n M_{\n{Xe}}}{\left({m_n+M_{\n{Xe}}}\right)^2}
	\left({1 - \cos \theta}\right)
\end{equation}
where $E_n$ is the energy of the incoming neutron, $m_n$ and $M_{\n{Xe}}$ are the masses of the neutron and
Xe nucleus, respectively, and $\theta$ is the scattering angle. The spread in measured recoil energies
mostly comes from the energy spread of the neutron source and the angular acceptance of the LXe and neutron
detectors due to their finite sizes.

Since the start of the XENON dark matter project, our group has already performed two direct measurements of
$\mathcal{L}_{\text{eff}}$. In this article we report results from a new measurement of this quantity, with an
improved apparatus and elaborated control of systematic uncertainties. These are the most accurate
measurements of $\mathcal{L}_{\text{eff}}$ in LXe to-date, down to 3~keV recoil energy.

\section{Experimental Setup}
\label{sec:setup}

\begin{figure}[!htb]
\begin{center}\includegraphics[width=1.0\columnwidth]{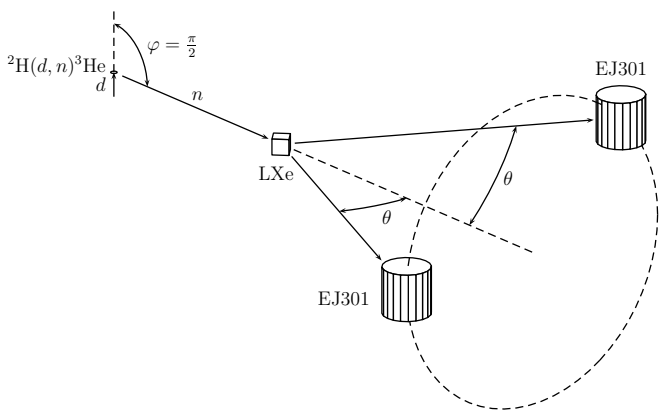}
\caption{Schematic of the experimental setup. A sealed-tube neutron generator, where deuterons of energy $E_d$
are incident upon a titanium deuteride target, produces neutrons at various angles $\varphi$. Some of the
neutrons emitted at an angle $\varphi = \frac{\pi}{2}$ scatter in the LXe detector at an angle $\theta$ and
are tagged by two EJ301 organic liquid scintillators neutron detectors.}
\label{fig:setup}
\end{center}
\end{figure}

% setup
The measurement of $\mathcal{L}_{\n{eff}}$ was carried out by irradiating a LXe detector with approximately
monoenergetic neutrons produced by a sealed-tube neutron generator \footnote{Neutron generator provided by
Schlumberger Princeton Technology Center, 20 Wallace Road, Princeton Jct., New Jersey 08550} and placing two
organic liquid scintillator detectors
% todo~\cite{eljentechnologies}
at different azimuthally symmetric positions
with respect to the neutron generator-LXe detector axis, see Fig.~\ref{fig:setup}.

\begin{figure}[!htb]
\begin{center}
	\includegraphics[width=1.0\columnwidth]{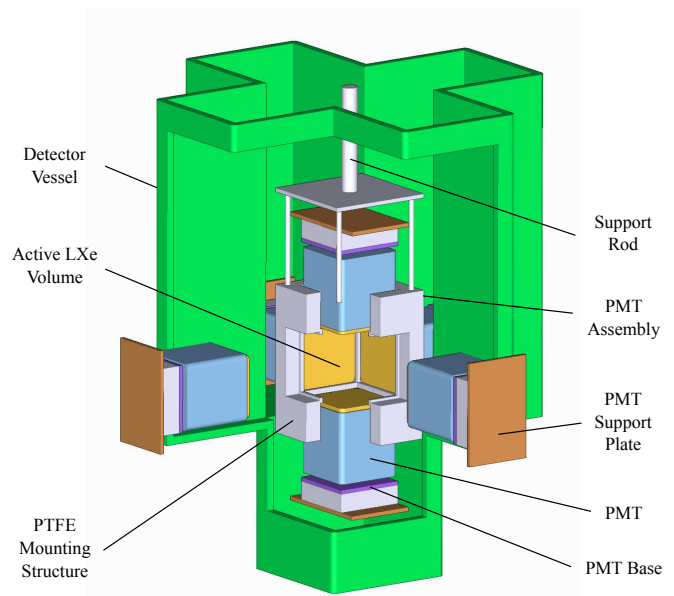}
\caption{Explosion drawing of the LXe detector.}\label{fig:xecube2}
\end{center}
\end{figure}

% detector
The emphasis for the design of the LXe detector has been placed on the reduction of non-active LXe and other
materials in the immediate vicinity of the active LXe volume, and on the maximization of the scintillation
light detection efficiency. The active LXe volume is a cube with sides of length $2.6 \1{cm}$ viewed by six
$2.5 \1{cm} \times 2.5 \1{cm}$ Hamamatsu R8520-406 Sel photomultiplier tubes (PMTs), the same type as used in
the XENON100 experiment~\cite{Aprile:2010um} but selected for high quantum efficiency (QE). The PMTs have a
special Bialkali photocathode for low temperature operation down to $-110^\circ\1{C}$ and have an average room
temperature QE of 32\% \footnote{Measured QE values for all six PMTs provided by Hamamatsu.} at 178~nm,
the wavelength at which Xe scintillates. The PMTs are operated with positive high voltage bias so that the
PMT metal body and photocathode are at ground potential. This ensures that the active volume remains at zero
electric field. This is a prerequisite to this measurement, since $\mathcal{L}_{\n{eff}}$
is defined as the relative scintillation efficiency of nuclear recoils
at zero field. A polytetrafluoroethylene (PTFE) frame serves as a mounting structure and alignment guide for
the PMTs so that each PMT window covers a side of the cubic active volume.  The PMT assembly is housed in a
stainless steel detector vessel, surrounded by a vacuum cryostat. The detector vessel has a special cross
shape that closely follows the contours of the PMT assembly to minimize the probability that neutrons scatter
before or after an interaction in the active volume.  The PTFE mounting structure is suspended from the top by
a stainless steel rod fixed to a linear displacement motion feedthrough. The motion feedthrough allows the
adjustment of the vertical position of the assembly from the outside. A schematic of the detector
is shown in Fig.~\ref{fig:xecube2}.

\begin{figure}[!htb]
\begin{center}
	\includegraphics[width=1.0\columnwidth]{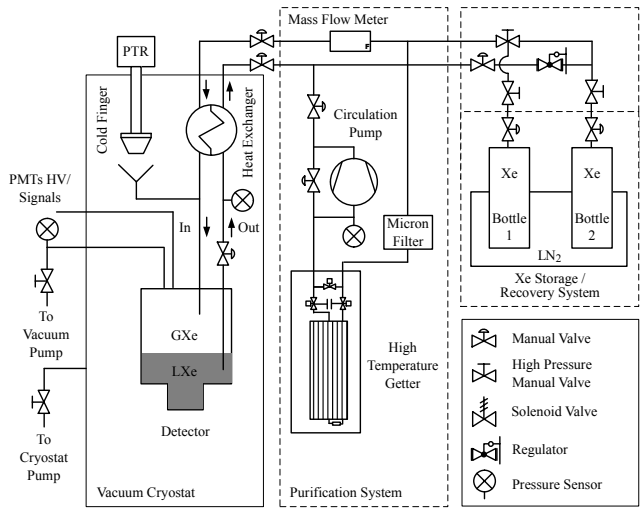}
\caption{Xenon gas system used for continuous purification of the LXe.}\label{fig:gas}
\end{center}
\end{figure}

% cooling, recirculation, etc
A schematic of the gas handling, liquefaction and re-circulation system developed for this experiment is shown
in Fig.~\ref{fig:gas}. The Xe is purified in the gas phase by circulating it through a hot getter
%todo ~\cite{saes}
and re-liquified efficiently using a heat exchanger~\cite{Giboni:2011wx}. The LXe temperature is kept constant
with an Iwatani PDC08 pulse tube refrigerator (PTR) delivering 24 W of cooling power at 165~K with an
air-cooled 1.5~kW helium compressor. Thanks to the PTR stability and low maintenance, the cooling system
developed for this detector has enabled us to acquire data over a two month long, uninterrupted run.  During
the measurements, the LXe temperature was maintained at $-94^\circ\1{C}$ which corresponds to a vapor pressure
of $2\1{atm}$. Further details on the design and performance of the cooling system developed for this
experiment are presented in Ref.~\cite{Giboni:2011wx}.

% neutron generator
Neutrons with an average energy of 2.5~MeV were produced via the
$^2\n{H}\!\left({d,n}\right)\!\vphantom{He}^3\n{He}$ reaction in a compact sealed-tube neutron generator
provided by Schlumberger.
%todo ~\cite{schlumberger}.
The generator was operated at deuteron energies of 60, 65, 75,
or 80~keV and with deuterium beam currents ranging from 60~to $100 \1{\mu A}$. The target is a
self-regenerating titanium deuteride (TiD) thick target.

For non-relativistic deuterons ($E_d < 20 \1{MeV}$), the energy of neutrons
emitted in the $^2\n{H}\!\left({d,n}\right)\!\vphantom{He}^3\n{He}$ reaction is given by~\cite{Csikai:1987}:
\begin{widetext}
\begin{equation}
	E_n^{1/2} = \frac{\left({m_d m_n E_d}\right)^{1/2}}{m_{\n{He}} + m_n} \cos \varphi
	+ \frac{\left\{{m_d m_n E_d \cos^2 \varphi + \left({m_{\n{He}} + m_n}\right) \left[{m_{\n{He}} Q
	+ \left({m_{\n{He}}-m_d}\right) E_d}\right]}\right\}^{1/2}}{m_{\n{He}} + m_n}
\end{equation}
\end{widetext}
where $m_d$ and $m_{\n{He}}$ are the deuteron and $^3\n{He}$ nucleus masses, respectively, $Q$ is the
$Q$-value of the reaction, and $\varphi$ is the neutron emission angle. At small and large emission angles the
neutron energy depends significantly on the deuteron energy. However, there is a minimum in both $\partial
E_n/\partial \varphi$ and $\partial E_n/\partial E_d$ at $\varphi \sim 100^\circ$, and consequently the energy
spread of neutrons produced is minimal near this angle. For this reason, the neutron generator was operated in
a configuration where deuterons are accelerated vertically and where the neutrons incident on the LXe detector
are those produced at $\varphi = \frac{\pi}{2}$ (see Fig.~\ref{fig:setup}).

\begin{figure}[!htb]
\begin{center}
	\includegraphics[width=1.0\columnwidth]{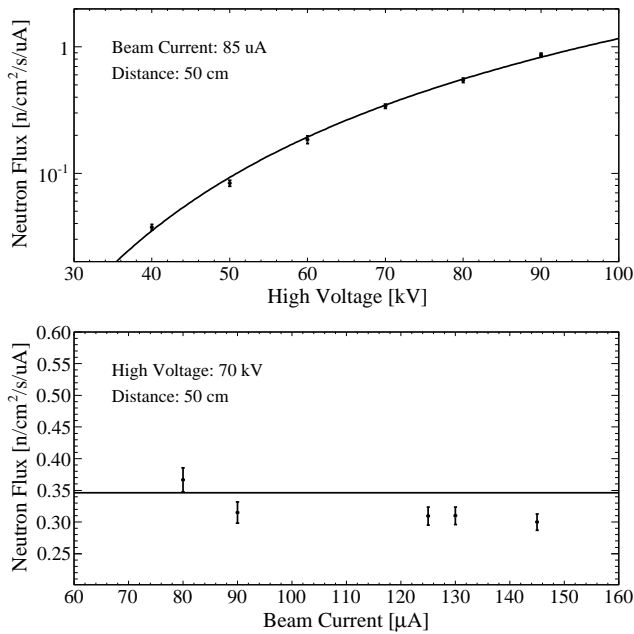}
\caption{Measurement (points) and theoretical calculation (solid line) of the neutron generator flux as a
function of high voltage (top) and beam current (down).}
\label{fig:minitron_flux}
\end{center}
\end{figure}

The expected neutron yield from the generator can be computed from its operating conditions, specifically the
deuteron energy, $E_d$, the deuterium beam current, and the monoatomic/diatomic deuterium beam composition.
Since deuterons are stopped in the target, the neutron angular yield $Y_{\n{tot}}\!\left({E_d,
\varphi}\right)$ is calculated from
\begin{equation}
	Y_{\n{tot}}\!\left({E_d, \varphi}\right) = x Y\!\left({E_d, \varphi}\right)
	+ 2 y Y\!\left({E_d/2, \varphi}\right)
\end{equation}
with
\begin{equation}
	Y\!\left({E_d, \varphi}\right) = \frac{\phi \, n_d}{\rho} \int_0^{E_d}
	\!\!\sigma\!\left({E'_d, \varphi}\right)
	\left[{\left({\frac{dE}{dx}}\right)_{\!\!d}\!\left({E'_d}\right)}\right]^{-1} dE'_d
\end{equation}
and the neutron energy distribution at angle $\varphi$ from
\begin{multline}
	N_{\n{tot}}\!\left({E_n, \varphi}\right) dE_n = \bigl[{x N\!\left({E_d, \varphi}\right) dE_d}\bigr. \\
	\bigl.{+ 2y N\!\left({E_d/2, \varphi}\right) dE_d}\bigr] \frac{dE_n}{dE_d}
\end{multline}
with
\begin{equation}
	N\!\left({E_d, \varphi}\right) dE_d = \frac{\phi \, n_d}{\rho}
	\sigma\!\left({E_d, \varphi}\right)
	\left[{\left({\frac{dE}{dx}}\right)_{\!\!d}\!\left({E_d}\right)}\right]^{-1} dE_d
\label{eq:energy_angle}
\end{equation}
where $\phi$ is the incident deuteron flux, $n_d$ the number density of deuterium atoms in the target and
$\rho$ its density, $\sigma\!\left({E_d, \varphi}\right)$ is the
$^2\n{H}\!\left({d,n}\right)\!\vphantom{He}^3\n{He}$ differential cross-section, $\left({dE/dx}\right)_d$ the
deuteron stopping power of the target, and $x$/$y$ the monoatomic/diatomic beam fractions.
%todo ~\cite{some reference}
The $^2\n{H}\!\left({d,n}\right)\!\vphantom{He}^3\n{He}$ neutron production cross-section is taken from the
ENDF/B-VII.0 database~\cite{endf-vii}, the stopping power for protons from the PSTAR
database~\footnote{http://physics.nist.gov/PhysRefData/Star/Text/PSTAR.html}, and the monoatomic/diatomic
deuterium beam fraction is taken as 0.05/0.95 \footnote{Andrew Bazarko (personal communication, 2010)}. The
result of the computation is shown in Fig.~\ref{fig:minitron_flux}.

Measurements of
the neutron flux were also carried out using a Nuclear Research Corporation NP-2 portable neutron monitor. The
NP-2 neutron monitor is a boron trifluoride proportional counter surrounded by a low density polyethylene
cylinder to moderate the neutron flux before capture by boron. Its accuracy is $\pm 10\%$. The result of the
computation show remarkable agreement with the measurement as a function of applied high voltage (maximum
deuteron energy). The agreement with the measurement as a function of beam current is also good but worsens as
the beam current increases. This effect is expected since, as the measured beam current increases, a larger
fraction of it can be due to electronic leakage current instead of deuterium current. The energy spread
of neutrons is assumed to be dominated by the energy loss of deuterons in the target
and by neutrons produced at different angles that scatter back into the LXe detector direction.

% liquid scintillators, alignment
The organic liquid scintillator neutron detectors used in this measurement are 3~inch diameter Eljen
Technologies M510 detector assemblies filled with EJ301 \footnote{EJ301 is the commercial name used for
$\n{C}_6\n{H}_4(\n{CH}_3)_2$. It is indentical to the more commonly known proprietary names of NE213 and
BC501A.}. The liquid scintillator is encapsulated in an aluminium cylindrical container, coupled to a 3~inch
ET Enterprises 9821B PMT. EJ301 is especially adapted to fast neutron detection in the presence of $\gamma$
radiation due to its excellent pulse shape discrimination (PSD) characteristics.
%todo ~\cite{psd}.

The alignment of the LXe detector and the neutron generator was performed with an auto-levelling laser mounted
on a tripod. The horizontal line and the vertical line projected by the laser were used to set the height of
the neutron generator target to the center of the LXe detector and to mark the position of both
on the laboratory floor.
The laser was also used to align the two EJ301 neutron detectors with respect to the LXe detector. For each
angle, the distance between the LXe detector and the two EJ301 neutron detectors was chosen to produce a
recoil energy spectrum with a spread due to the angular acceptance of the cell of 10\% to 20\%. The
desired EJ301 neutron detector positions were marked on the floor using 1.5~m aluminium rules,
while their
height was set with the help of the laser and a vertical rule. The EJ301 neutron detectors were supported at
the desired positions around the LXe detector by their own laboratory stands.

The horizontal distance between the neutron generator and the LXe detector was fixed at $40 \1{cm}$.  The
distance from the LXe detector to the EJ301 liquid scintillators varied from $100 \1{cm}$, for the scattering
angles corresponding to low-energy recoils, to $40\1{cm}$, for the scattering angles corresponding to higher
energies. The positioning accuracy of the EJ301 neutron detectors is estimated to be better than $5
\1{mm}$.

% trigger, acquisition
The data acquisition electronics are largely the same as those of the XENON100 experiment. The signals from
the six LXe PMTs are fed into a Phillips 776 $\times 10$ amplifier with two amplified outputs per channel. The
first output of each channel is digitized by a 14-bit CAEN V1724 100~MS/s flash ADC with 40~MHz bandwidth
while the second output is fed to a Phillips 706 leading edge discriminator. The discriminator thresholds are
set at a level of -15~mV, which corresponds to 0.5~photoelectrons (pe). The logic signals of the six
discriminator outputs are added with a linear fan-in and discriminated to obtain a 2-fold PMT coincidence
condition. The 2-fold PMT coincidence logic signal is passed to a $10\, \mu\n{s}$ holdoff circuit to prevent
re-triggering on the tail of the LXe scintillation signal and constitutes the LXe trigger.

The signals from the two EJ301 PMTs are also fed into the $\times 10$~amplifier, with one output digitized by
the flash ADC unit, and the other output discriminated. However, since we do not expect a signal for both EJ301
PMTs in the same event, the two signals are multiplexed into a single channel before digitization. The two
discriminator outputs are delayed and also digitized to serve as a demultiplexing code. A copy of the
discriminator outputs is combined into an OR gate and forms the EJ301 trigger.

The neutron time-of-flight (TOF) is measured with an Ortec 566 time to amplitude converter (TAC), where the
LXe trigger is used as the ``start'' signal and a delayed copy of the EJ301 trigger as the ``stop'' signal.
The delay of the copy of the EJ301 trigger signal can be varied to calibrate the TOF measurement. The TAC
output signal is also digitized by the flash ADC unit and the TOF is computed by the event processing program.

\begin{figure}[!htb]
\begin{center}\includegraphics[width=1.0\columnwidth]{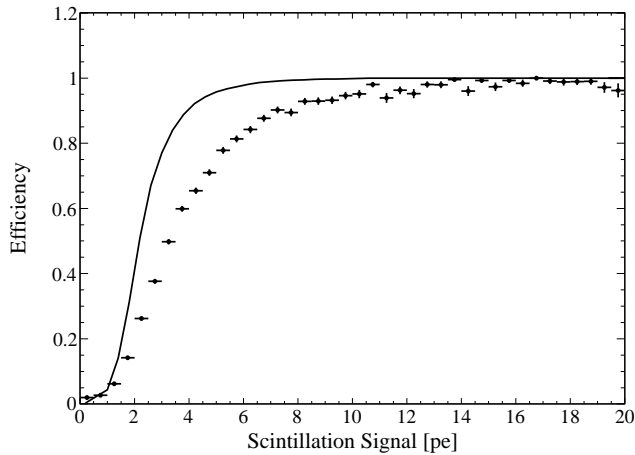}
\caption{Measured (points) and simulated (solid curve) efficiency of the 2-fold coincidence LXe
trigger. The measured trigger efficiency is used to extract $\mathcal{L}_{\n{eff}}$ (Sec.~\ref{sec:leff}) from
the neutron scattering measurements.}
\label{fig:trigger_efficiency}
\end{center}
\end{figure}

% trigger efficicency
Finally, for neutron scattering measurements, the trigger is taken as a coincidence within a 200~ns window of
the LXe trigger and the EJ301 trigger.
The low-energy roll-off of the trigger efficiency can contribute significantly~\cite{Manalaysay:2010mb} to the
systematic uncertainty on $\mathcal{L}_{\n{eff}}$, if it is not properly understood. As an improvement over
previous measurements of $\mathcal{L}_{\n{eff}}$, where the efficiency was based solely on simulations, the
efficiency of the trigger setup described above was measured in addition to its simulation. The efficiency
measurement was done using a $^{22}\n{Na}$ source (a $\beta^+$ emitter) placed between the LXe detector and a
sodium iodide NaI(Tl) detector. The back-to-back pair of 511~keV $\gamma$ rays from
the $\beta^+$ annihilation interact effectively at the same time in the LXe and the NaI(Tl) detectors. The
NaI(Tl) detector was positioned such that the solid angle it subtended at the source was larger than the one
subtended by the active LXe volume. This ensures that the whole active volume of the LXe detector is probed.
For this measurement, the triggering signal consisted of the discriminated signal of the output of the NaI(Tl)
detector. In addition to the signals of the LXe PMTs, the trigger signal of the normal LXe trigger was
digitized with the flash ADC. The efficiency is inferred by computing the fraction of events accompanied by a LXe
trigger signal as a function of their measured number of photoelectrons. Fig.~\ref{fig:trigger_efficiency}
shows the resulting trigger efficiency.

The expected efficiency has also been computed via a detailed Monte Carlo simulation. The simulation takes
into account the spatial distribution of the light collection efficiency within the active LXe volume, PMT
quantum efficiencies, PMT first dynode collection efficiencies, PMT gains, variations in the single
photoelectron pulse height distribution, and discriminator trigger thresholds. The result is also shown in
Fig.~\ref{fig:trigger_efficiency}. The discrepancy between the simulated and measured efficiencies is
attributed to the limited accuracy in modelling hardware components of the trigger. Consequently, the
measured trigger efficiency is used to extract $\mathcal{L}_{\n{eff}}$ (Sec.~\ref{sec:leff}) from the neutron
scattering measurements. In effect, this discrepancy only stresses the need to perform a measurement of the trigger
efficiency instead of relying solely on simulation.

\section{Data Analysis}

\subsection{Calibration}

% pmt gains
The LXe PMT gains were measured under single photoelectron conditions with a pulsed blue light emitting diode
(LED), embedded in the PTFE mounting structure. The gains were equalized to a mean value of $2.0\times 10^6$
by adjusting the individual PMT anode bias voltages at the beginning of the experiment and were monitored
regularly throughout. At the operating conditions chosen, the single photoelectron resolution of the LXe PMTs
has a mean value of $60\%$.

\begin{figure}[!htb]
\begin{center}
	\includegraphics[width=1.0\columnwidth]{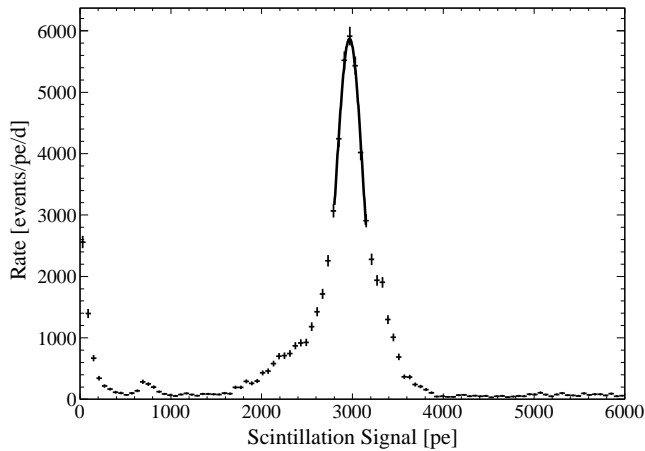}
	\caption{Scintillation light spectrum of the $100\1{\mu Ci}$ $^{57}\n{Co}$ source used to calibrate the
	LXe light yield, $L_y$. The peak at $\sim \!3000\1{pe}$ is the 122~keV photoelectric absorption peak.
	Also visible is the Xe $30\1{keV}$ characteristic X-ray at $\sim\!700\1{pe}$. This calibration gives
	a scintillation light yield of $L_y = 24.3 \1{pe/keV}$.}
	\label{fig:light_yield}
\end{center}
\end{figure}

% light yield
The electronic recoil energy scale is calibrated using 122~keV $\gamma$ rays since $\mathcal{L}_{\n{eff}}$ is
defined as the scintillation light yield of nuclear recoils relative to the yield of $\gamma$ rays of that
energy. Calibrations with an external $100\1{\mu Ci}$ $^{57}\n{Co}$ source were taken regularly throughout the
duration of the experiment. Fig.~\ref{fig:light_yield} shows the scintillation spectrum of a calibration with
the $^{57}\n{Co}$ source. The scintillation light yield was measured to be $L_y = 24.14 \pm 0.09 \n{(stat)}
\pm 0.44 \n{(sys)} \1{pe/keVee}$ with a resolution $\left({\sigma/E}\right)$ of 5\%. This very high light
yield, combined with the 90\% trigger efficiency at 7~pe, implies that energy spectra do not suffer from
efficiency losses down to energies as low as 0.3~keVee.

Since the attenuation length of 122~keV $\gamma$ rays in LXe is $3\1{mm}$~\cite{xcom} the external
$^{57}\n{Co}$ source mostly probes the light yield in the outer layers of the active volume. However, since
the elastic scattering mean free path of 2.5~MeV neutrons in LXe is $\sim\!20\1{cm}$, the expected
spatial distribution of nuclear recoils is uniform. Hence, the quantity of interest is the average light
yield over the whole volume. The spatial uniformity of the light collection has been investigated using a
light propagation simulation that takes into account the detailed geometry of the active volume, the
photocathode coverage of the PMTs, and the reflectivity of PTFE. The relative light yield variation over the
active volume was calculated to be less than 2\%, with a maximum variation of 5\% near the edge.

% tof calibration
The TOF calibration was performed with the $^{22}\n{Na}$ source placed between the LXe detector and the EJ301
neutron detectors. The two emitted 511~keV $\gamma$ rays emitted interact in both detectors and give a $\n{TOF} = 0$
calibration point. The time delay between the LXe and EJ301 triggers was varied with a delay generator over a
range of 32~ns to obtain the TOF calibration.

\subsection{Data Processing, Selection}

% processing
Events consist of the six LXe PMT waveforms, the waveform of the triggered EJ301 neutron detector, as well as
the waveform of the TAC output. For each event, the processing software searches the LXe PMT waveforms for
scintillation signals and records several parameters for each pulse found: position, area, height, etc.  The
same procedure is applied to the EJ301 neutron detector waveform. In addition, a PSD parameter is defined as
the fraction of the total EJ301 scintillation signal contained within the tail of the pulse. The tail of the
pulse is defined as a region with a lower boundary at 30~ns after the pulse peak and an upper
boundary at the time where the pulse returns to 1\% of the peak amplitude. Finally, the TOF is computed from
the height of the TAC ouput and the TOF calibration values. For a LXe detector and EJ301 neutron detector
distance of 1~m, the the typical neutron TOF is 45~ns.

% data selection, cuts
Three selection cuts are applied to the data: a PSD cut to select neutron interactions in the EJ301 neutron
detectors, a lower energy threshold cut on the EJ301 scintillation signal, and a TOF cut.

\begin{figure}[!htb]
\begin{center}
    \includegraphics[width=1.0\columnwidth]{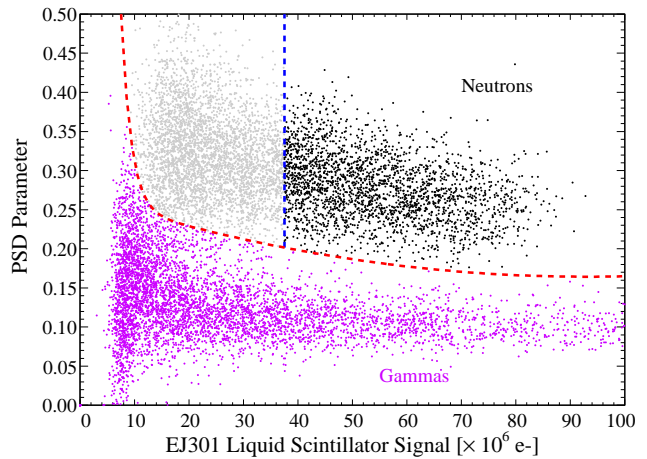}
	\caption{Neutron interactions (gray/black) and gamma interactions (violet) for one of the two EJ301
	neutron detectors. The neutron events selected by the PSD cut (red dashed line) and the EJ301 energy
	threshold cut (blue dashed line) are shown in black.}\label{fig:psd}
\end{center}
\end{figure}

% psd
The PSD cut uses the $\gamma$/n discrimination capabilities of the EJ301 liquid scintillator and allows the
selection of neutron interactions in the EJ301 with high efficiency. However, since this cut is based on the
scintillation signal, the discrimination power degrades with decreasing energy and it is preferable to apply
an energy threshold cut on the EJ301 scintillation signal. This reduces the background from
neutrons that scattered in other materials, in addition to the scatter in the LXe detector, since
the neutron energy was reduced and hence cannot deposit the maximum energy expected. Fig.~\ref{fig:psd} shows
the events selected by the combination of the PSD cut and the EJ301 energy threshold cut for one of the two
EJ301 neutron detectors.

% tof
A subtle effect can be observed in the TOF calibration data. Since the LXe trigger comes from the coincidence
of two LXe scintillation photons, the LXe scintillation light decay time systematically shifts the TOF
measurement of scintillation signals with fewer photons to lower TOF values. The effect is more pronounced for
$\gamma$ rays interacting in LXe due to the slower recombination time for electron recoils.
%todo ~\cite{Doke}.
The
TOF value is corrected for this effect in the event processing software. However, the correction does not
eliminate the spread in $\n{TOF}$ values due to this effect. To eliminate any possible bias, we chose TOF cuts
that contain the full neutron TOF peak for each scattering angle measurement, even if this results in a
higher contamination of the recoil spectrum by neutrons that scattered in other materials.

\subsection{Measured Recoil Distributions}

\begin{figure*}[htbp]
\begin{center}
	\begin{tabular}{c c}
		\includegraphics[width=0.9\columnwidth]{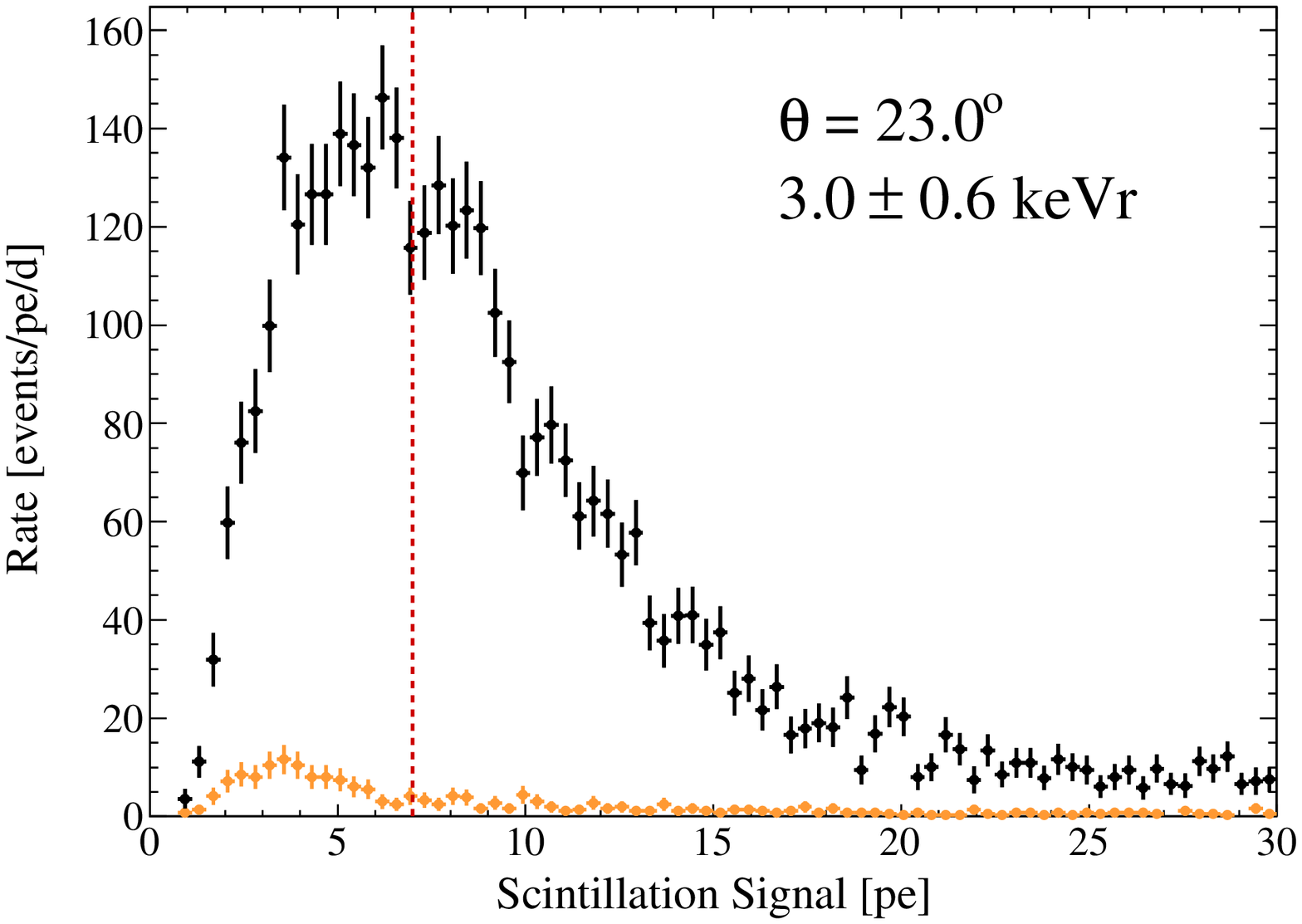} &
		\includegraphics[width=0.9\columnwidth]{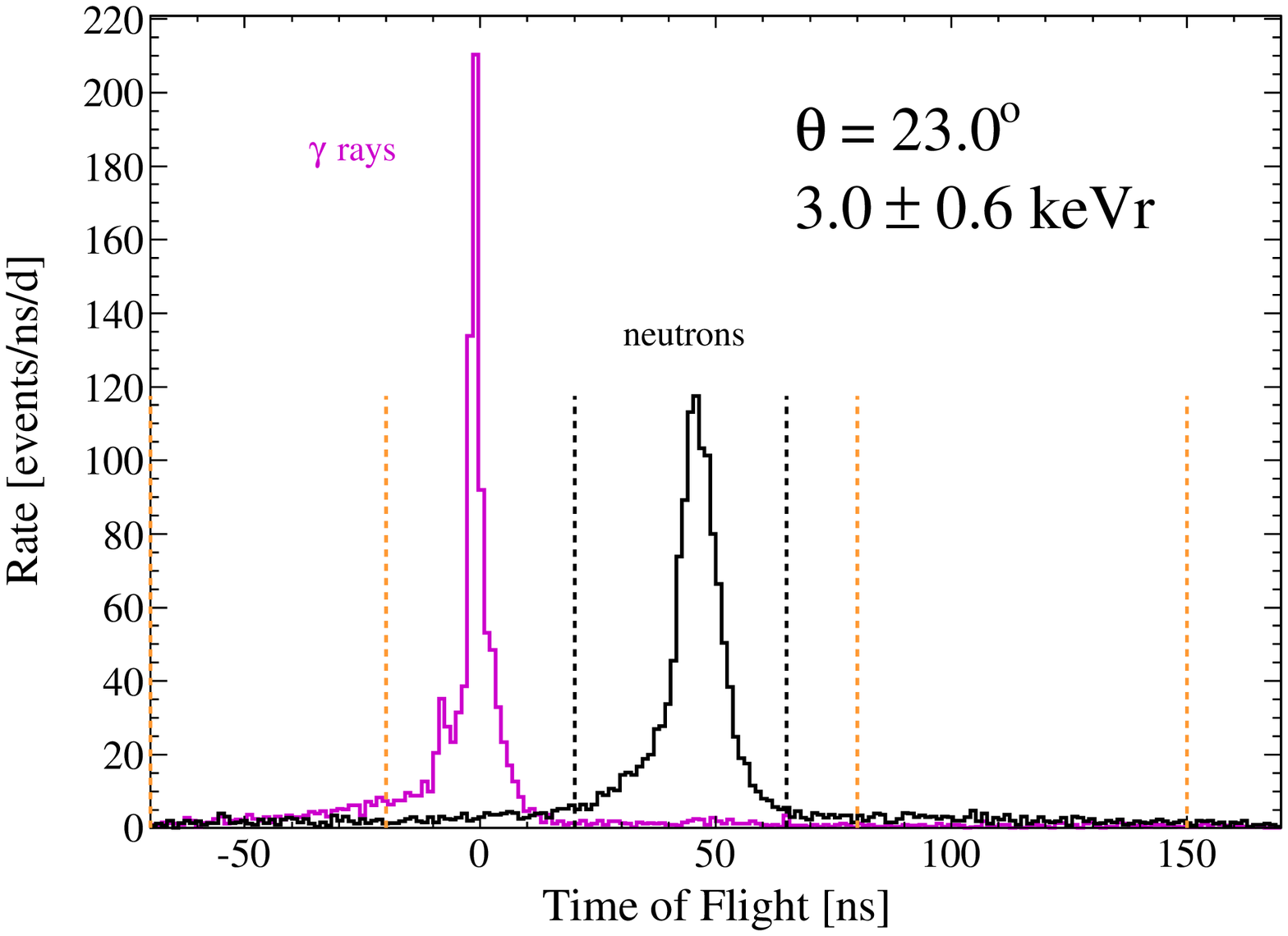} \\
		\includegraphics[width=0.9\columnwidth]{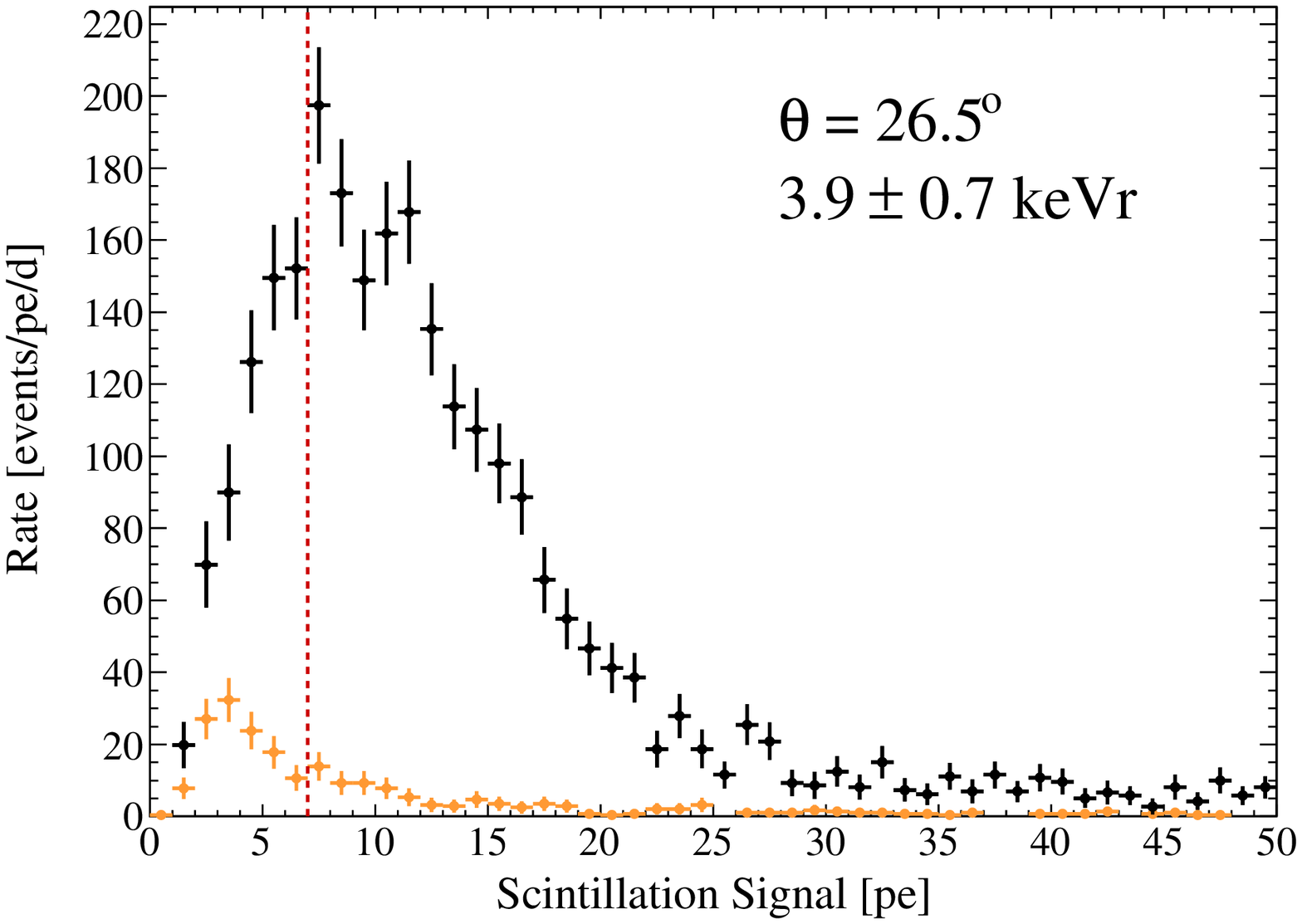} &
		\includegraphics[width=0.9\columnwidth]{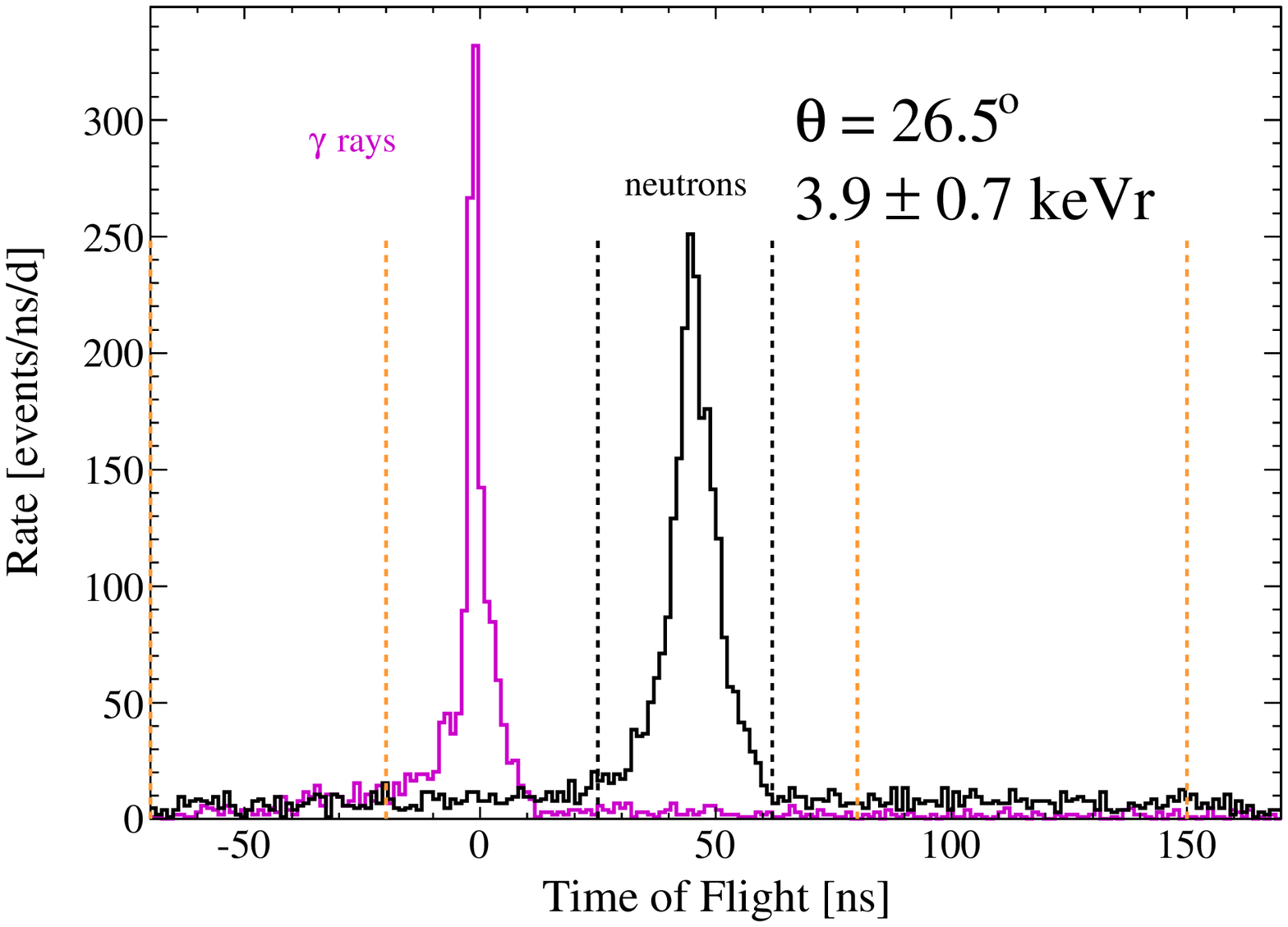} \\
		\includegraphics[width=0.9\columnwidth]{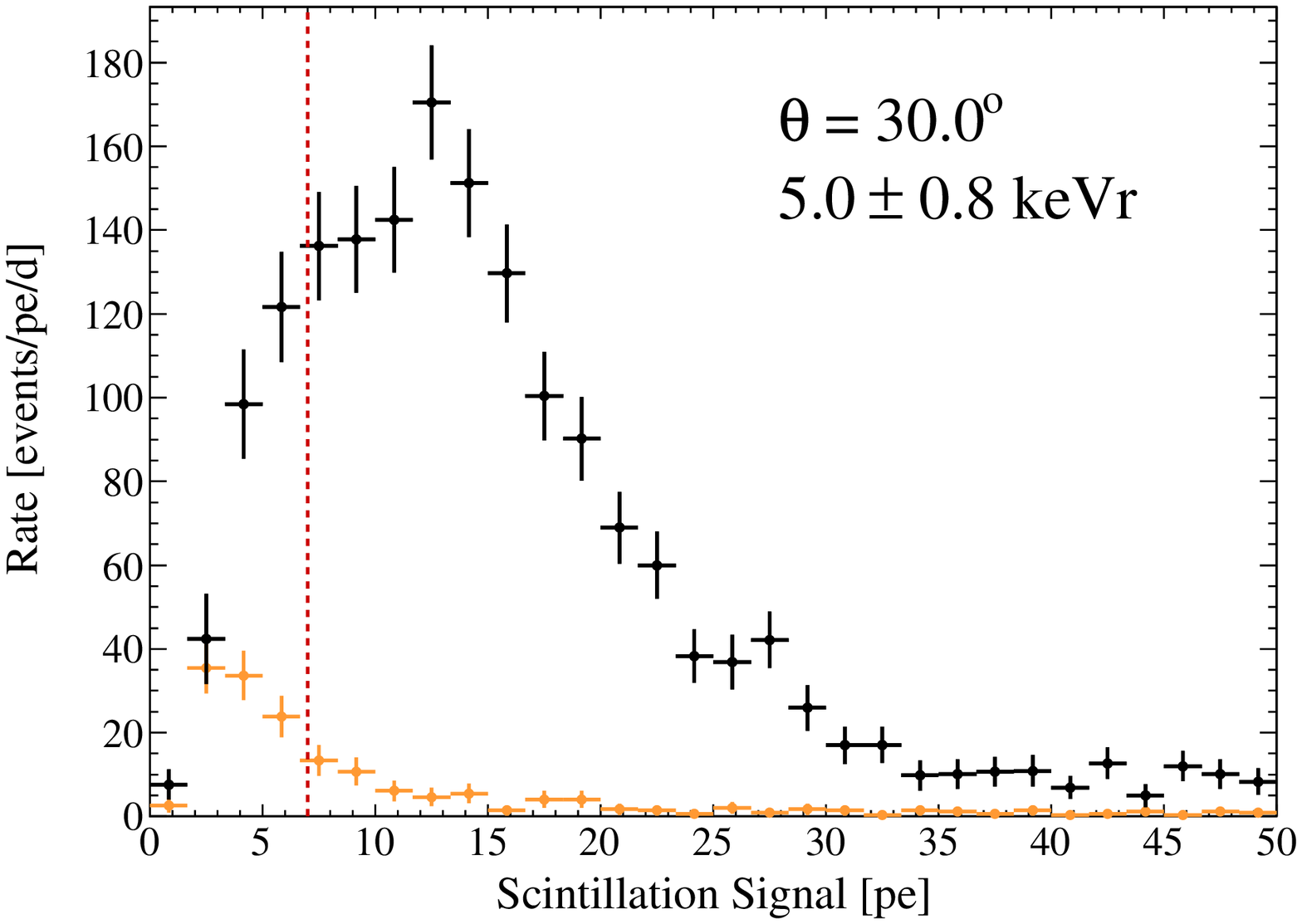} &
		\includegraphics[width=0.9\columnwidth]{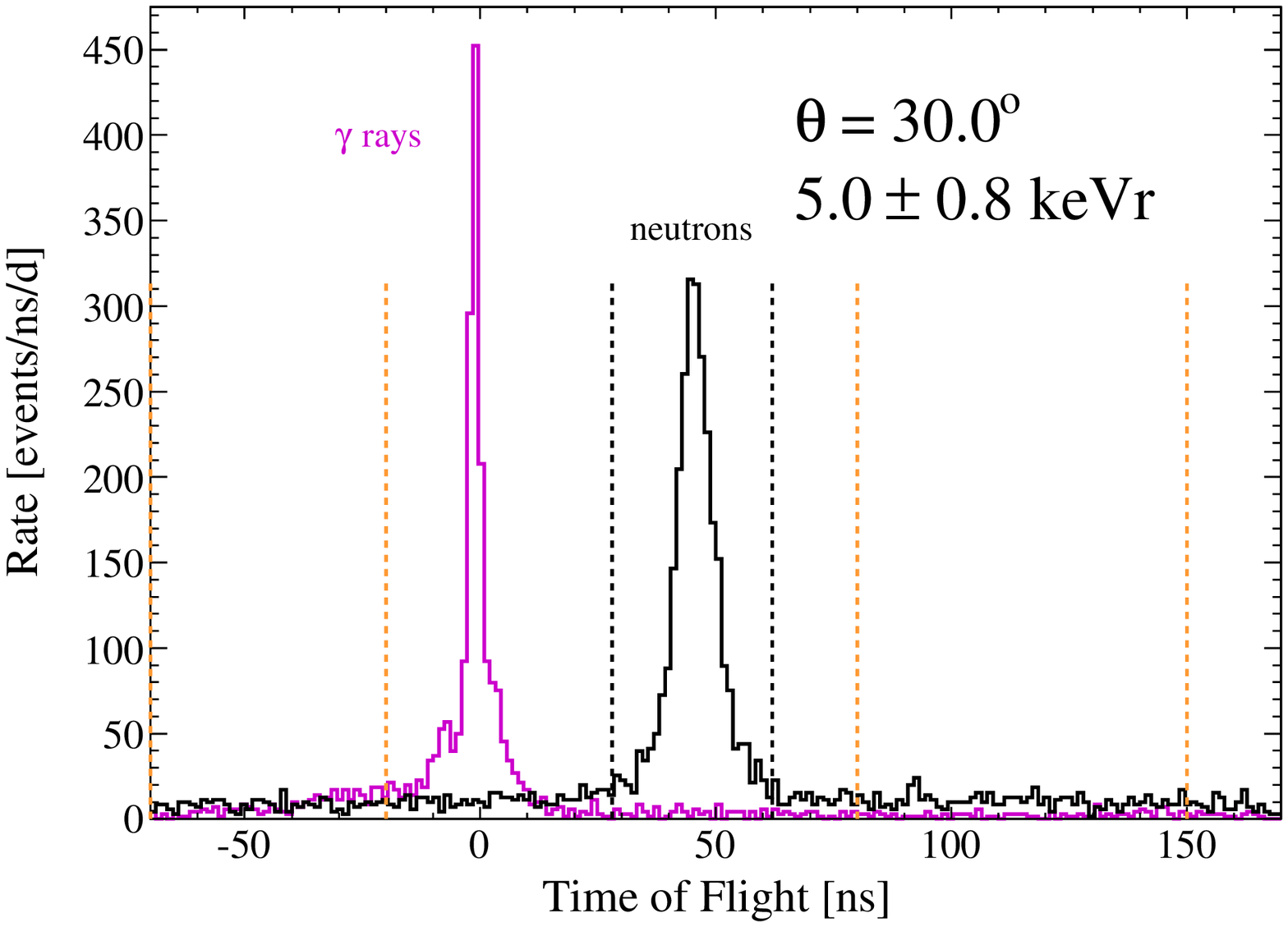} \\
		\includegraphics[width=0.9\columnwidth]{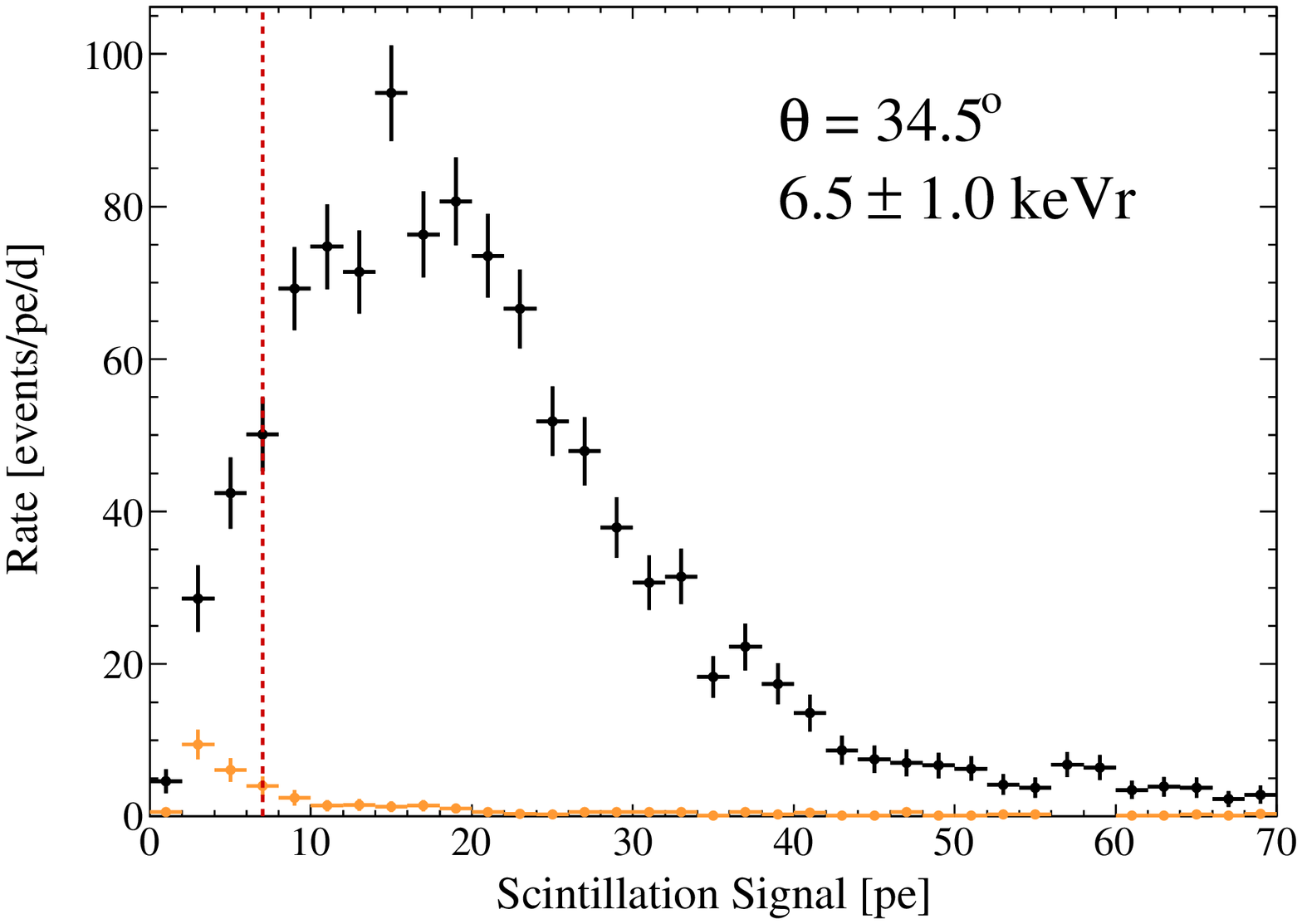} &
		\includegraphics[width=0.9\columnwidth]{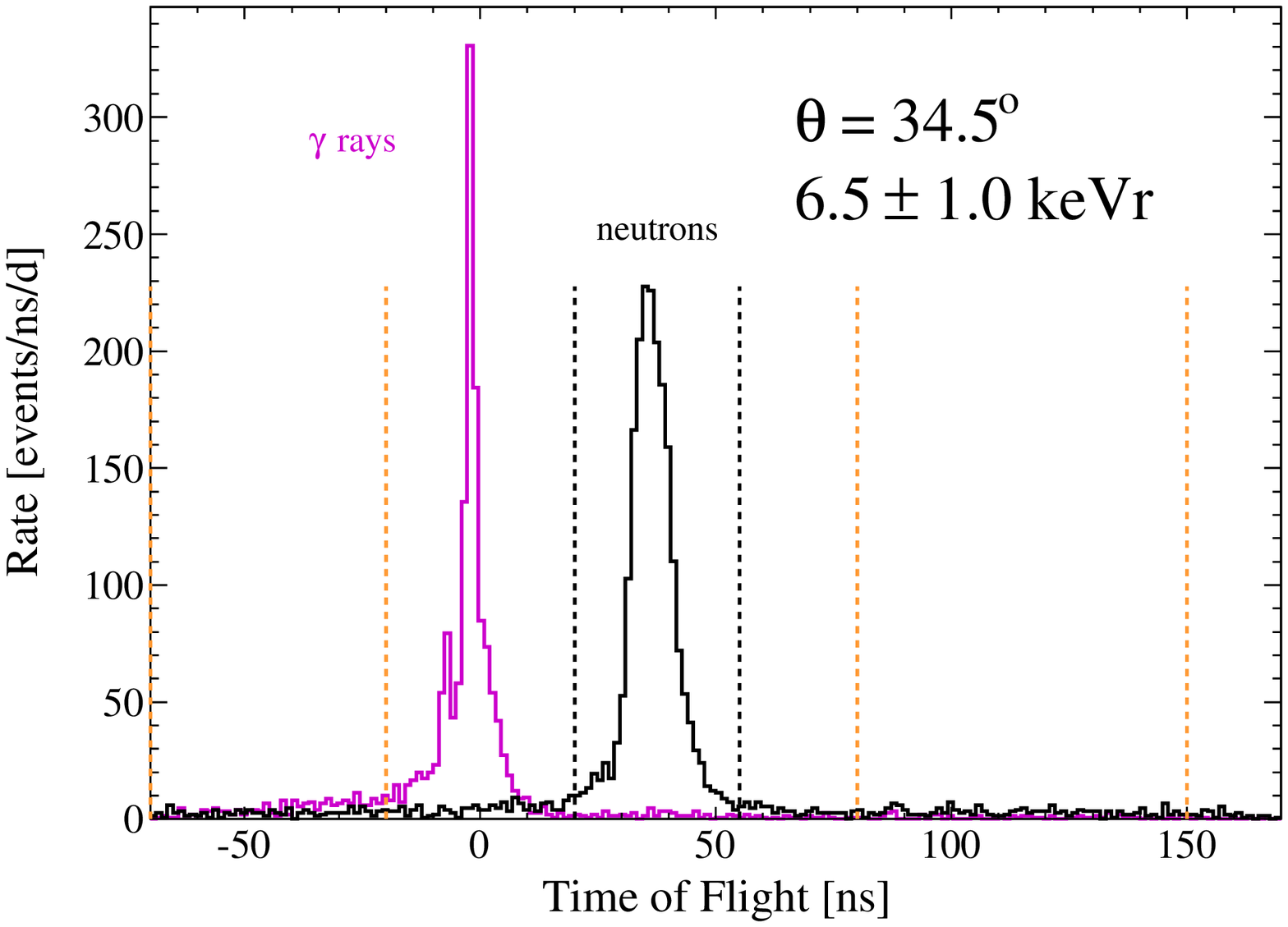} \\
	\end{tabular}
\caption{(left) Recoil energy spectrum for the $23^\circ$, $26.5^\circ$, $30^\circ$, and $34.5^\circ$
scattering angles, with PSD and EJ301 energy threshold cuts applied, for the TOF window indicated in the TOF
spectrum by the vertical dashed lines (right). The black histogram is the recoil energy spectrum, after
subtraction of the accidental spectrum shown as an orange histogram. As a reference, the 90\% measured trigger
efficiency is indicated by the vertical red dashed line. The accidental spectrum expectation is obtained from
the TOF windows before and after the main TOF peak, as indicated in the figure by the vertical orange dashed
lines. The accidental spectrum in the window before the peak is in agreement with the one after the peak. The
purple histogram is the TOF spectrum where the PSD cut is chosen to select $\gamma$ ray interactions in the
EJ301 neutron detectors.}
\label{fig:data1}
\end{center}
\end{figure*}

\begin{figure*}[htbp]
\begin{center}
	\begin{tabular}{c c}
		\includegraphics[width=0.9\columnwidth]{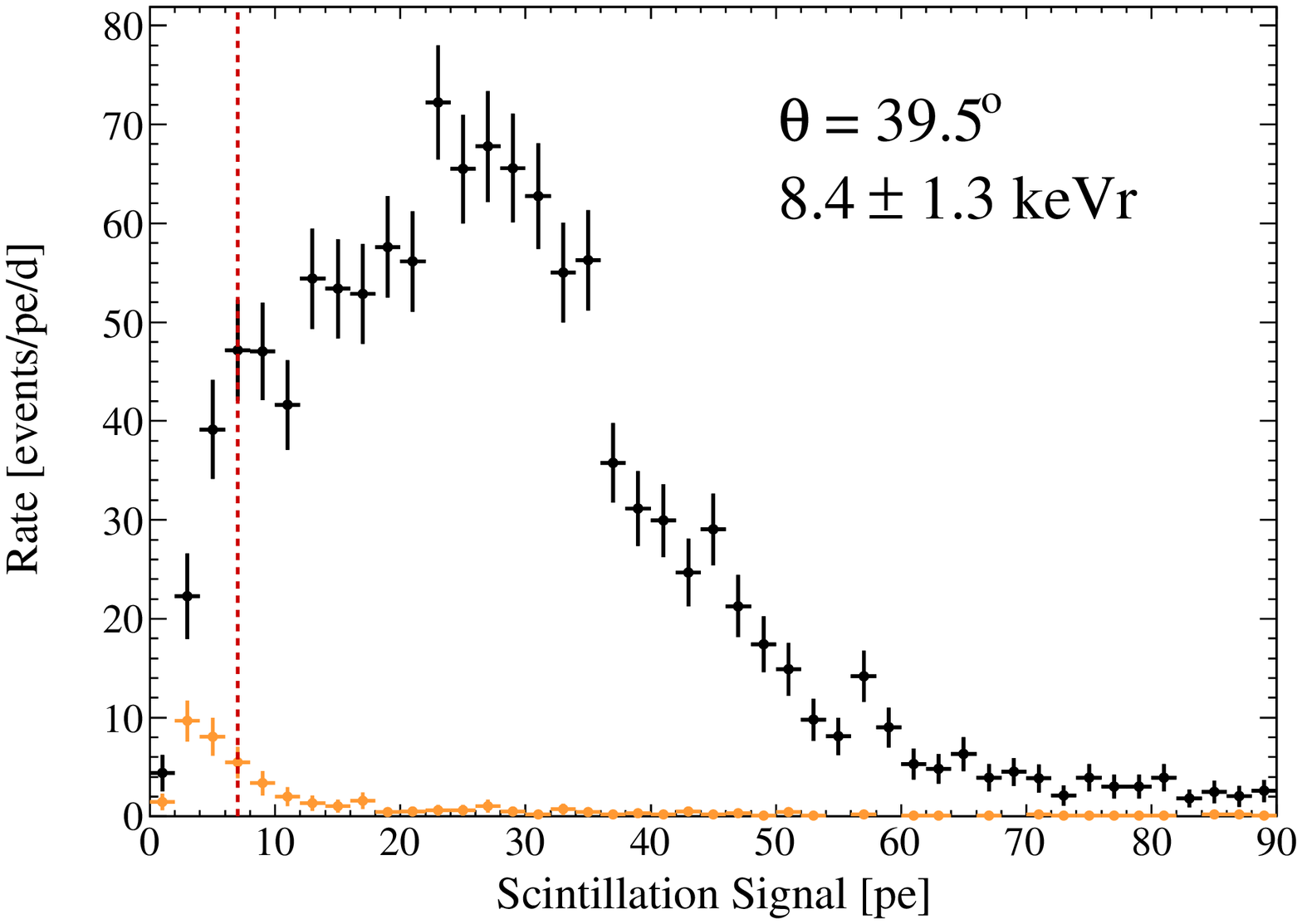} &
		\includegraphics[width=0.9\columnwidth]{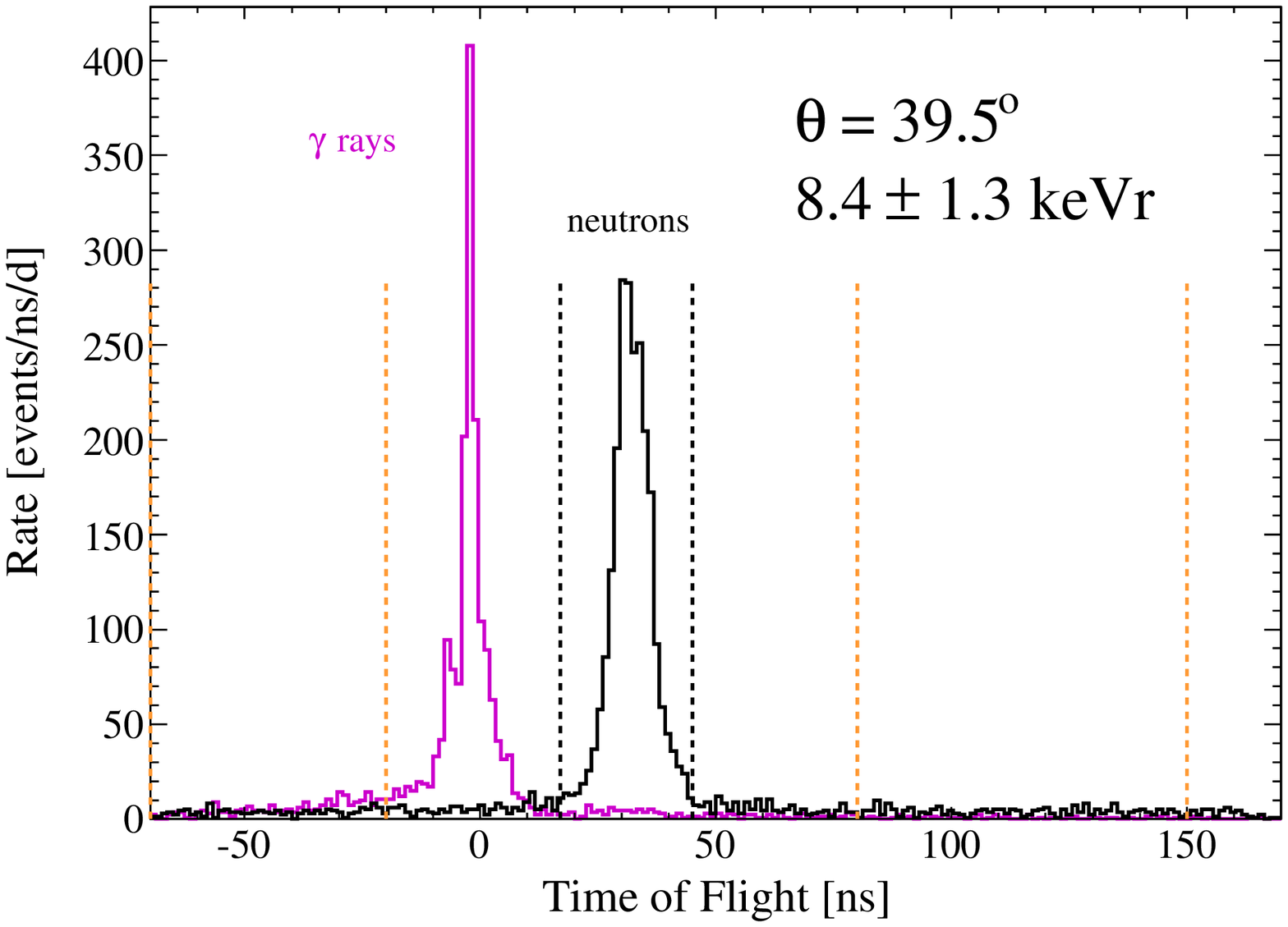} \\
		\includegraphics[width=0.9\columnwidth]{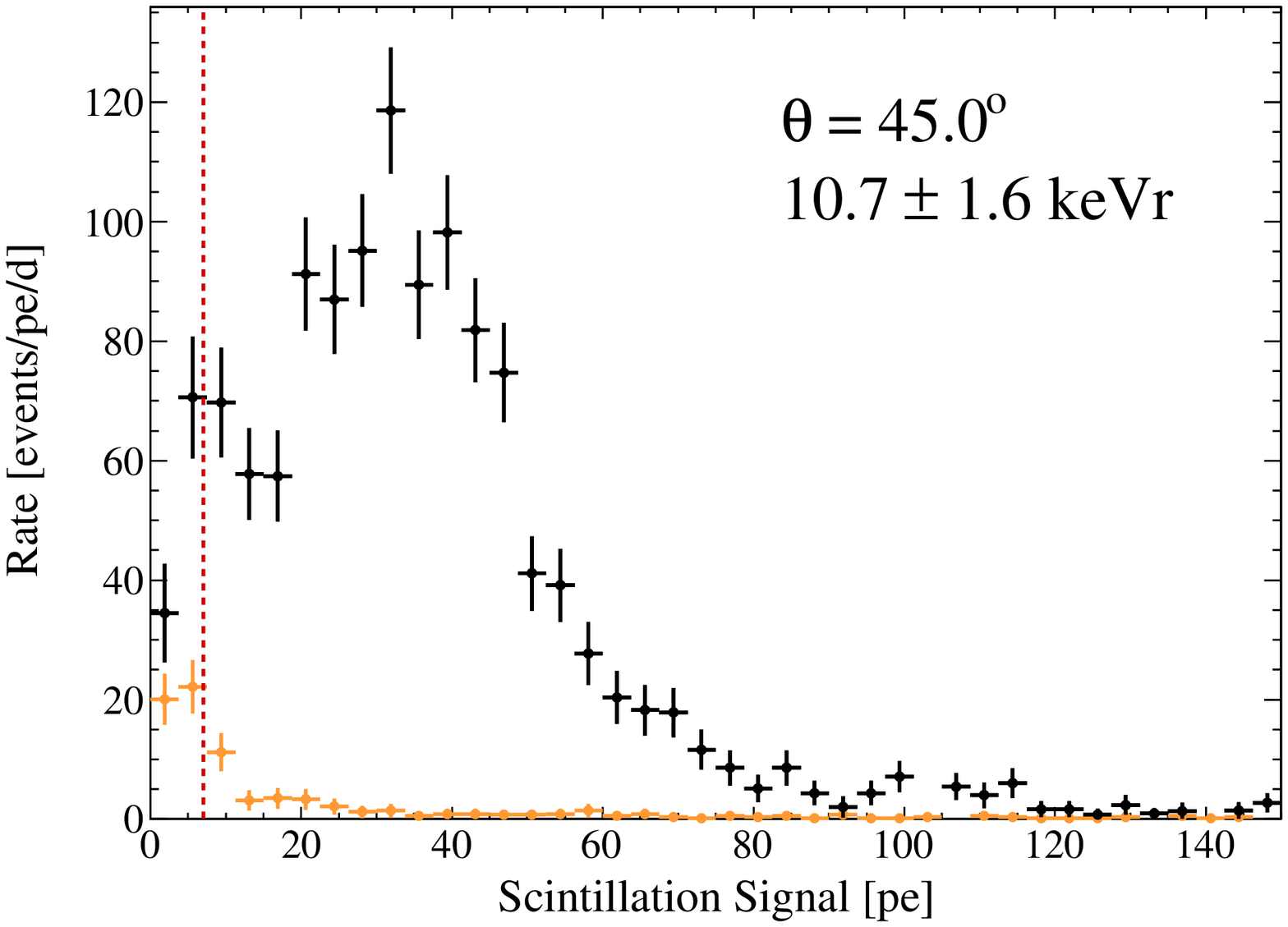} &
		\includegraphics[width=0.9\columnwidth]{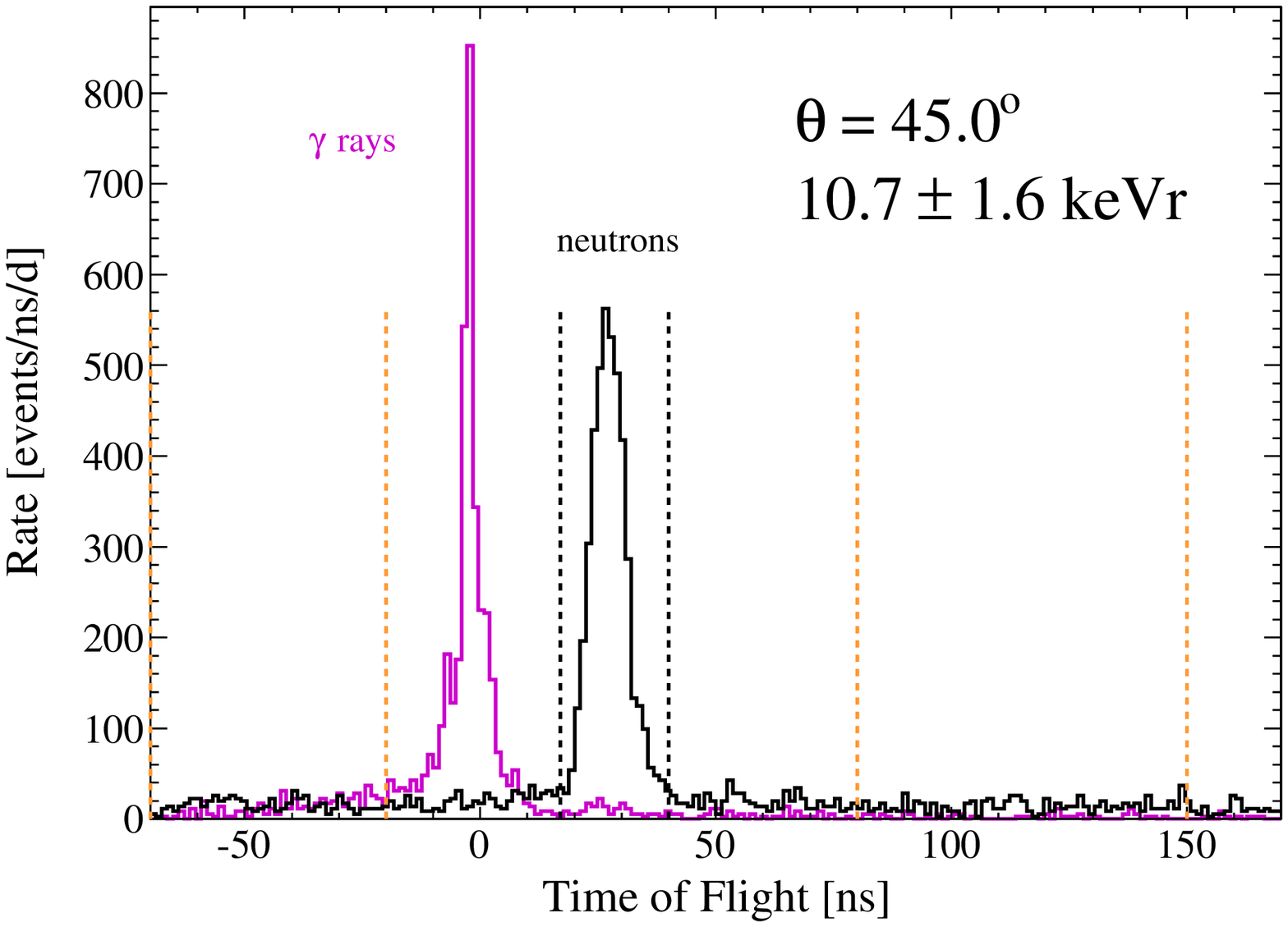} \\
		\includegraphics[width=0.9\columnwidth]{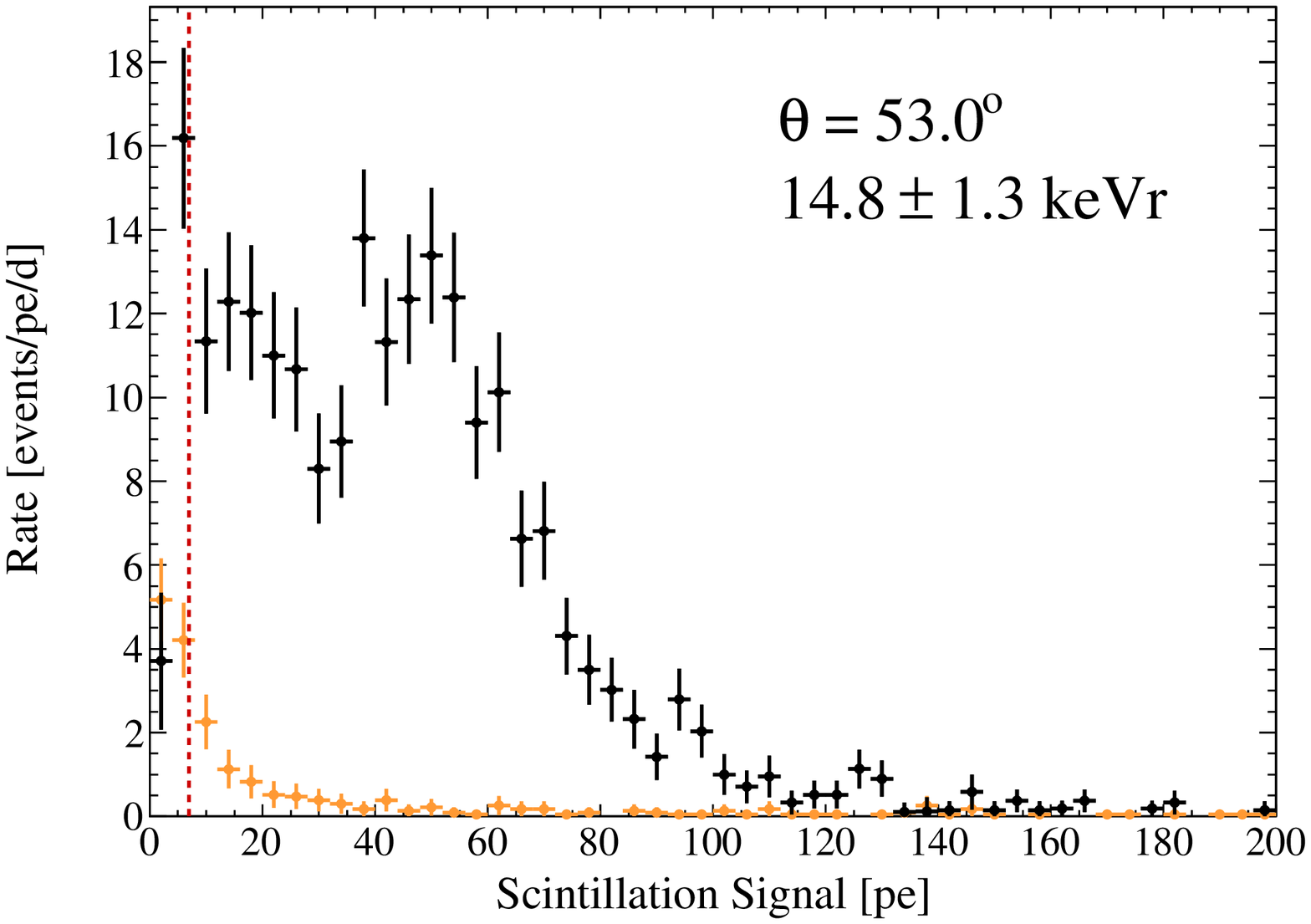} &
		\includegraphics[width=0.9\columnwidth]{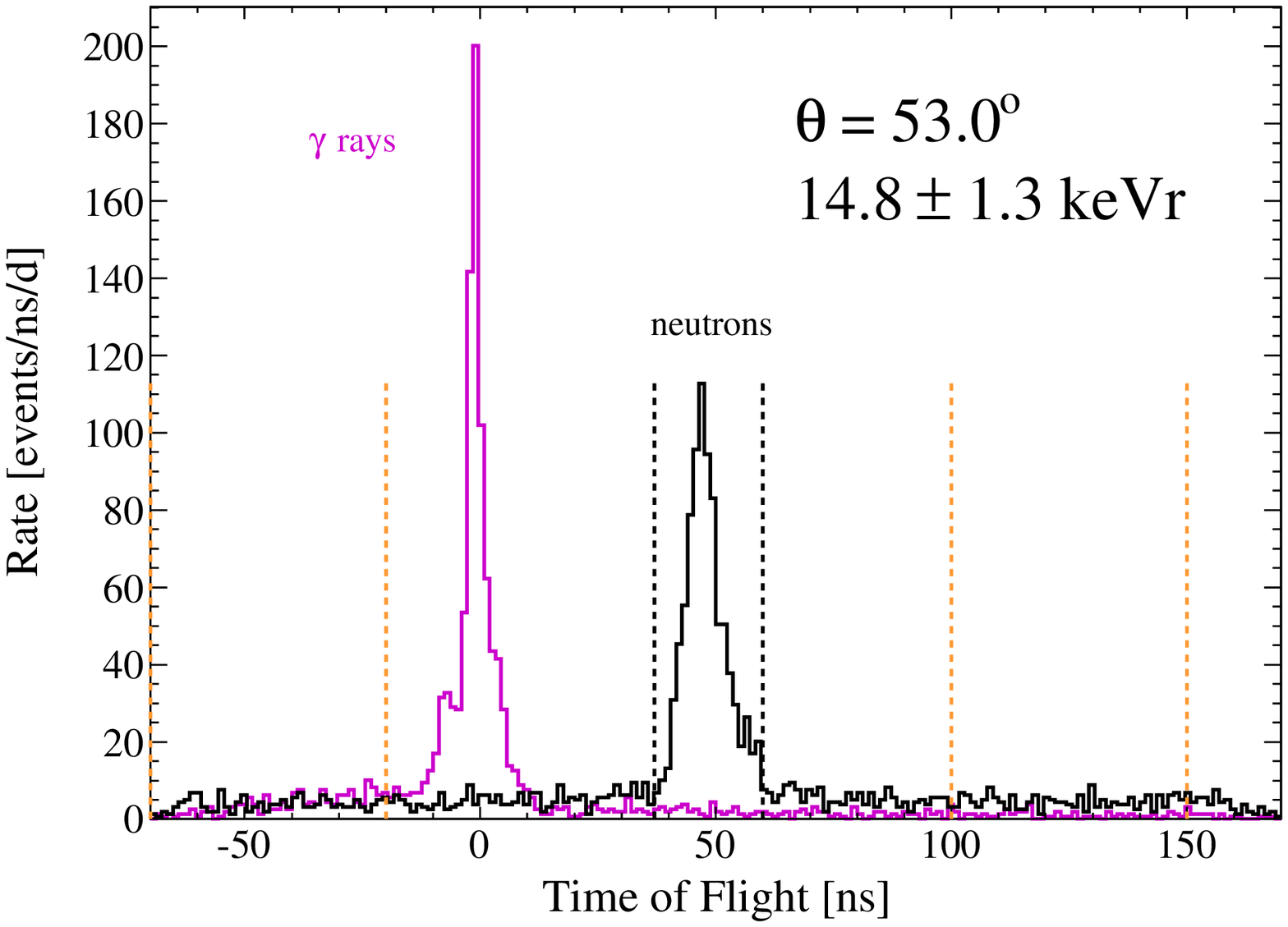} \\
		\includegraphics[width=0.9\columnwidth]{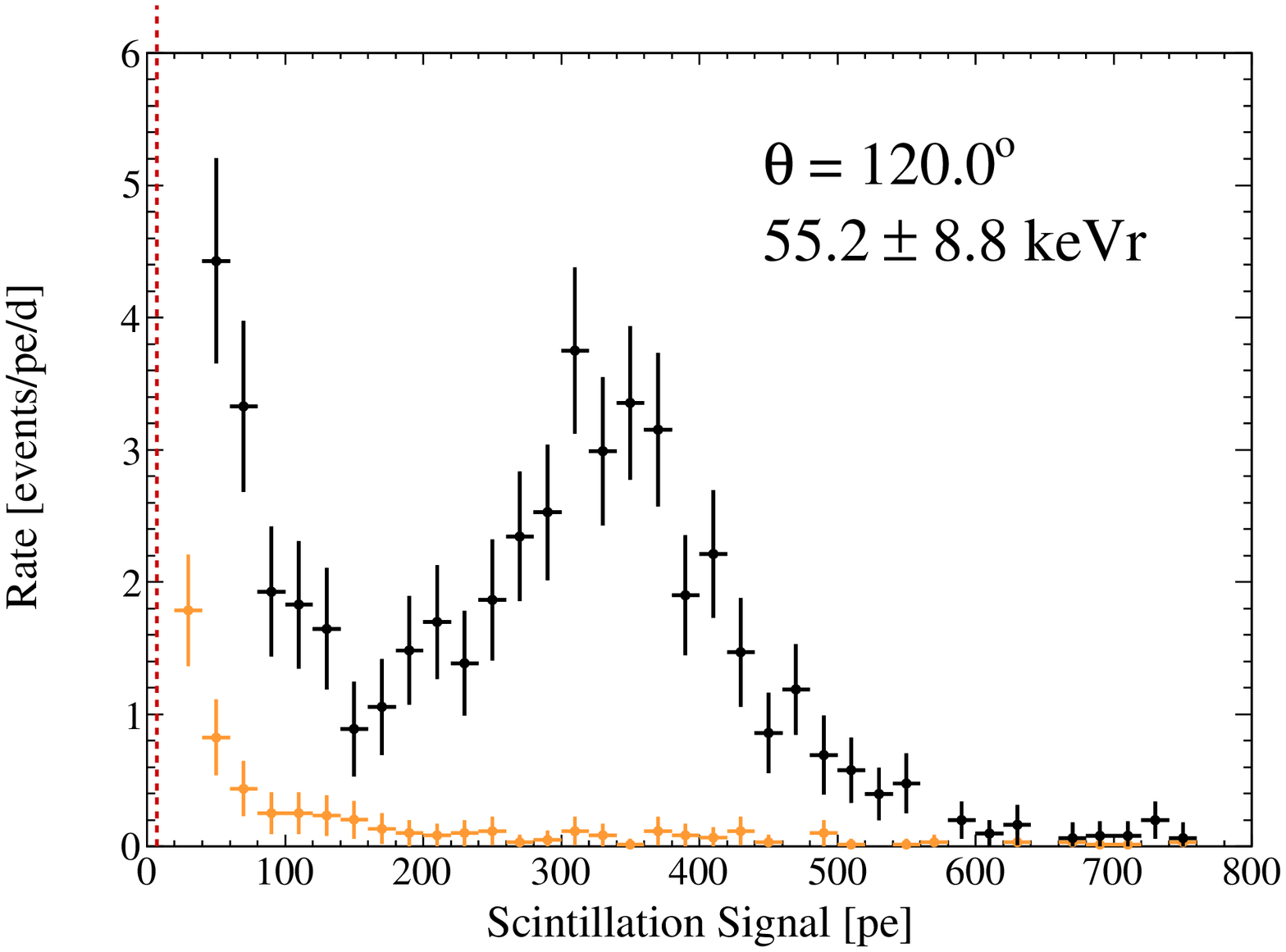} &
		\includegraphics[width=0.9\columnwidth]{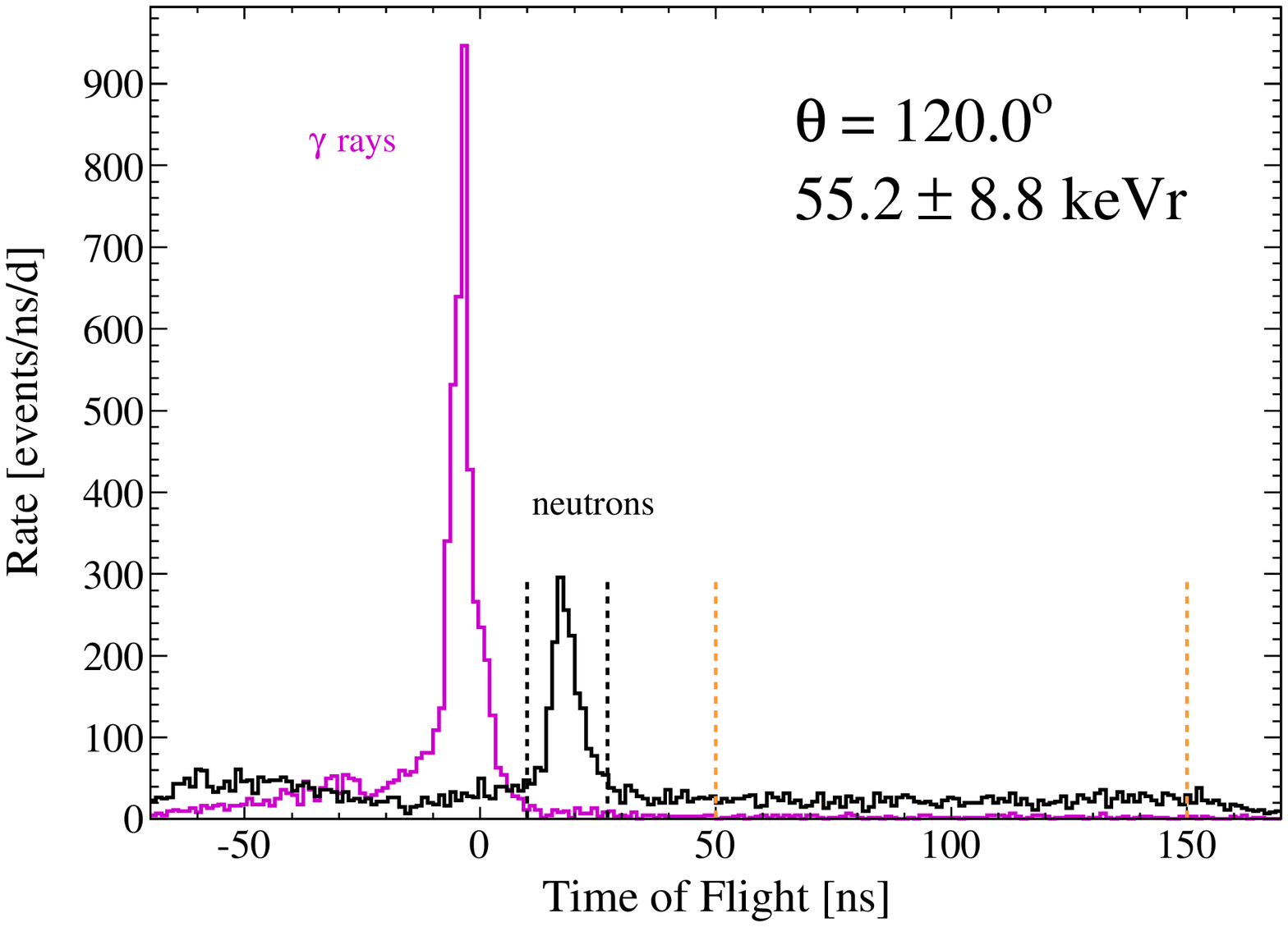} \\
	\end{tabular}
\caption{(left) Recoil energy spectrum for the $39.5^\circ$, $45^\circ$, $53^\circ$, and $120^\circ$
scattering angles, with PSD and EJ301 energy threshold cuts applied, for the TOF window indicated in the TOF
spectrum by the vertical dashed lines (right). The black histogram is the recoil energy spectrum, after
subtraction of the accidental spectrum shown as an orange histogram.  As a reference, the 90\% measured
trigger efficiency is indicated by the vertical red dashed line. The accidental spectrum expectation is
obtained from the TOF windows before and after the main TOF peak, as indicated in the figure by the vertical
orange dashed lines. The accidental spectrum in the window before the peak is in agreement with the one after
the peak. The purple histogram is the TOF spectrum where the PSD cut is chosen to select $\gamma$ ray
interactions in the EJ301 neutron detectors.}
\label{fig:data2}
\end{center}
\end{figure*}

% data
Neutron scattering data were acquired at eight different angles: $23^\circ$, $26.5^\circ$, $30^\circ$,
$34.5^\circ$, $39.5^\circ$, $45^\circ$, $53^\circ$, and $120^\circ$, corresponding respectively to recoil
energies $E_{\n{nr}}$ of $3.0\pm 0.6$, $3.9 \pm 0.7$, $5.0\pm 0.8$, $6.5 \pm 1.0$, $8.4 \pm 1.3$, $10.7 \pm
1.6$, $14.8 \pm 1.3$, and $55.2 \pm 8.8$ keV. The uncertainty in nuclear recoil energies, dominated by the
angular acceptance of the detectors, is extracted from the results of the Geant4 Monte Carlo simulation
described in Sec.~\ref{sec:mc}. For most angles, the first EJ301 neutron detector was placed in the plane of
the neutron generator and the LXe detector, while the second was placed at the same distance to the LXe
detector but higher above the laboratory floor,
with an azimuthal angle of about $45^\circ$. For all measurements the recoil
energy and TOF distributions of the two EJ301 neutron detectors are compatible.  Figures~\ref{fig:data1} and
\ref{fig:data2} show the measured recoil energy and TOF spectra for the all scattering angles.

All measurements show two very clear peaks in the TOF spectrum. The peak at $\n{TOF} = 0$ corresponds to
$\gamma$ rays that Compton scatter in the LXe detector before interacting in the EJ301 neutron detectors,
while the peak at later TOF values corresponds to neutrons. For the lower recoil energies the effect of the
spread of the TOF distribution to smaller values, discussed in the previous section, is noticeable
in Figures~\ref{fig:data1} and \ref{fig:data2}.

For energies of 6.5~keV and above, the peak in the recoil spectrum is clearly above the beginning of the
low-energy trigger efficiency roll-off. For these energies, $\mathcal{L}_{\n{eff}}$ could even be computed
directly without much uncertainty, simply by fitting a Gaussian to the peak. For energies below 6.5~keV, a
more sophisticated procedure that takes into account the trigger efficiency is warranted. The procedure used
to extract $\mathcal{L}_{\n{eff}}$ from the measured recoil energy spectra is detailed in Sec.~\ref{sec:mc}.

\subsection{Monte Carlo Simulation}
\label{sec:mc}

% simulation
Extensive Geant4
% todo ~\cite{geant4}
simulations of the expected neutron scattering rate, nuclear recoil energy
distribution and neutron TOF distribution were performed for each scattering angle. Each
simulation takes into account a realistic description of the neutron generator, LXe detector, detector vessel,
vacuum cryostat, support frame, as well as the measured positions of the two EJ301 neutron detectors. The
livetime of each scattering angle simulation is calculated from the expected neutron yield of the generator at
the operating conditions of the measurement and the results are scaled accordingly.

The information recorded in the simulation includes the energy, position, time, type of particle, and physical
process responsible for each energy deposit in the LXe detector, as well as the energy, time, and type of particle
for each energy deposit in the EJ301 neutron detectors. The energy and direction of the primary neutron is
sampled from the calculated energy-angle distribution of neutrons produced by the generator, as given by
eq.~\ref{eq:energy_angle}.

\begin{figure*}[htbp]
\begin{center}
	\begin{tabular}{c c}
		\includegraphics[width=0.9\columnwidth]{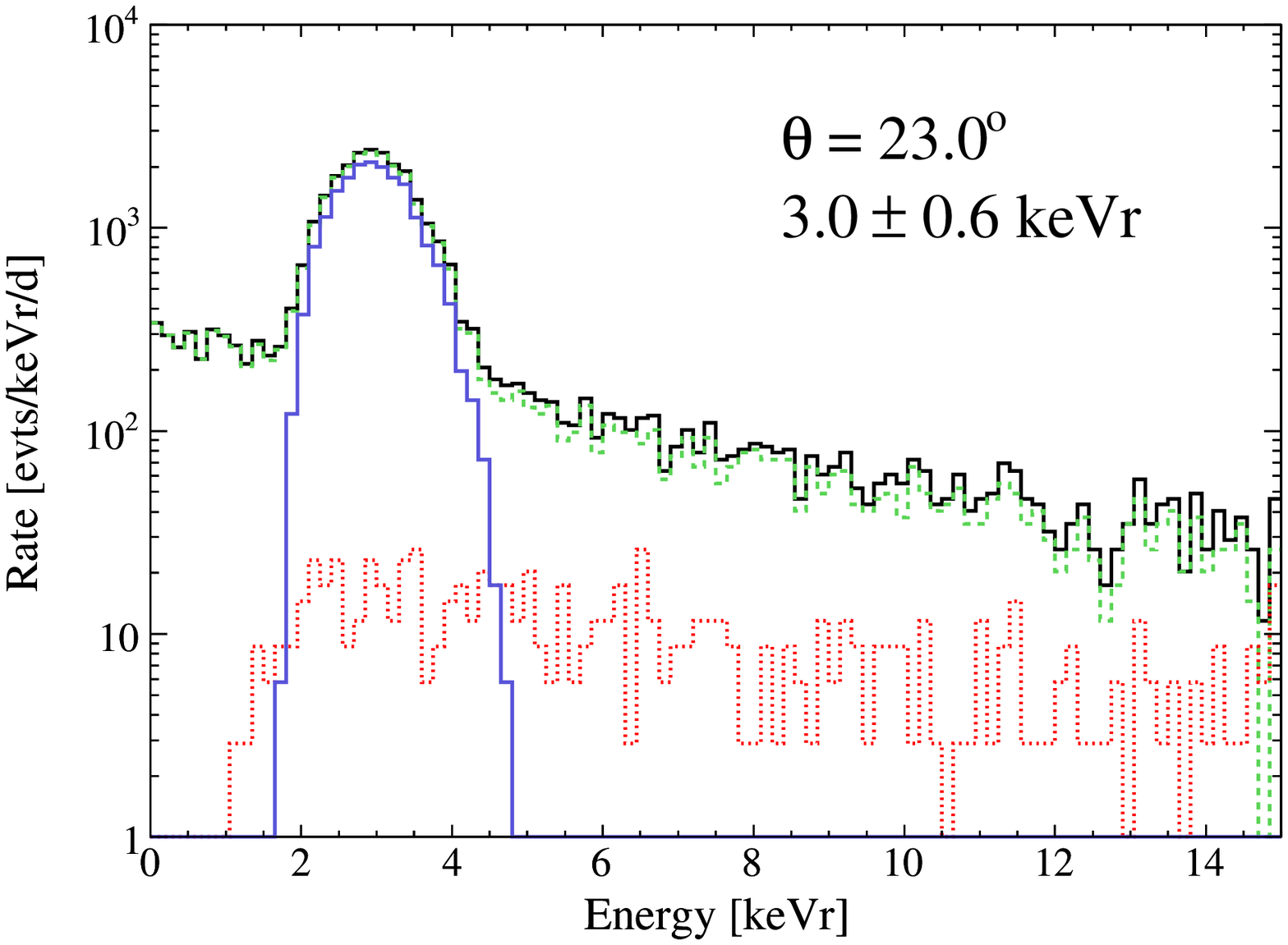} &
		\includegraphics[width=0.9\columnwidth]{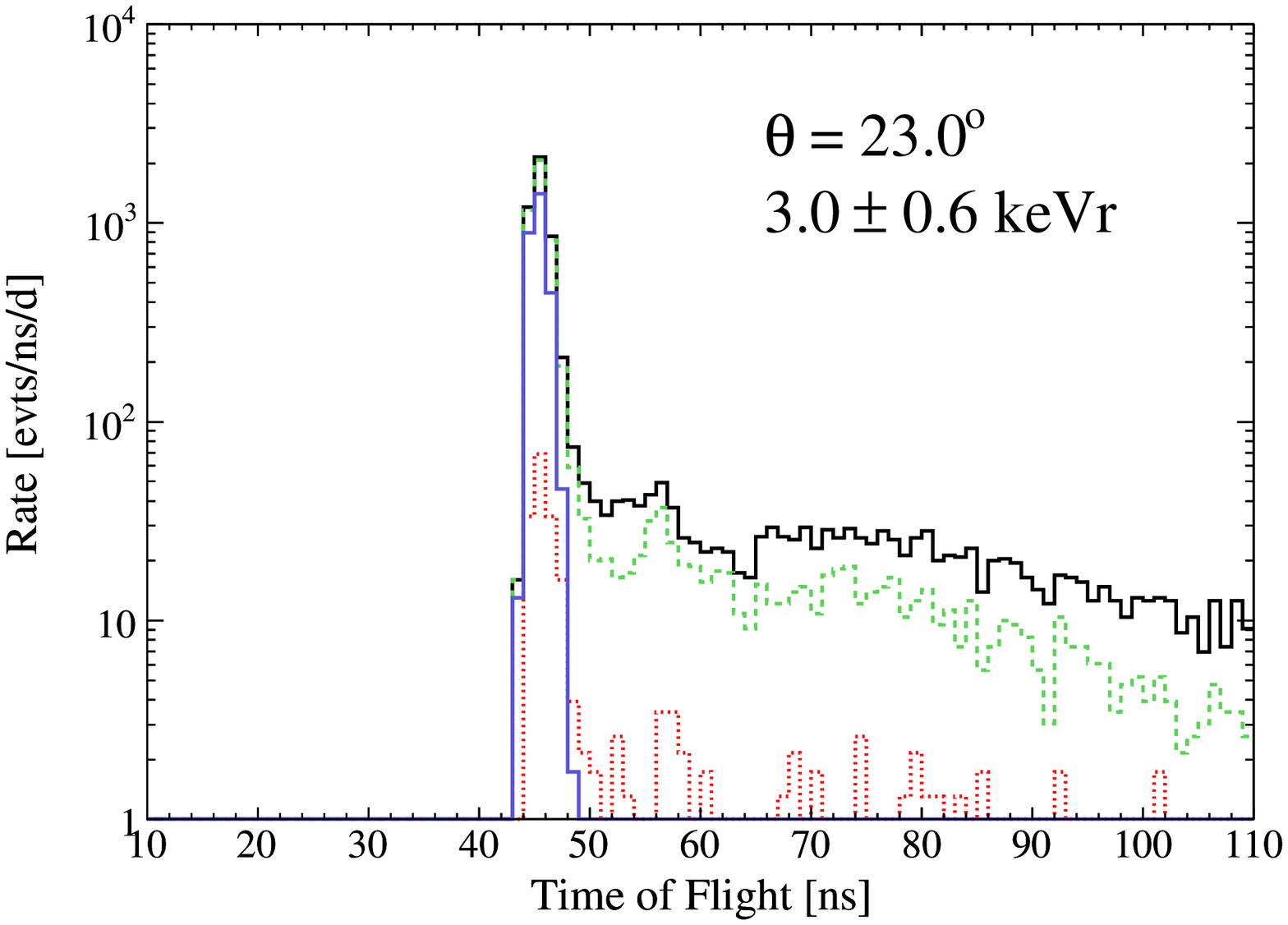} \\
		\includegraphics[width=0.9\columnwidth]{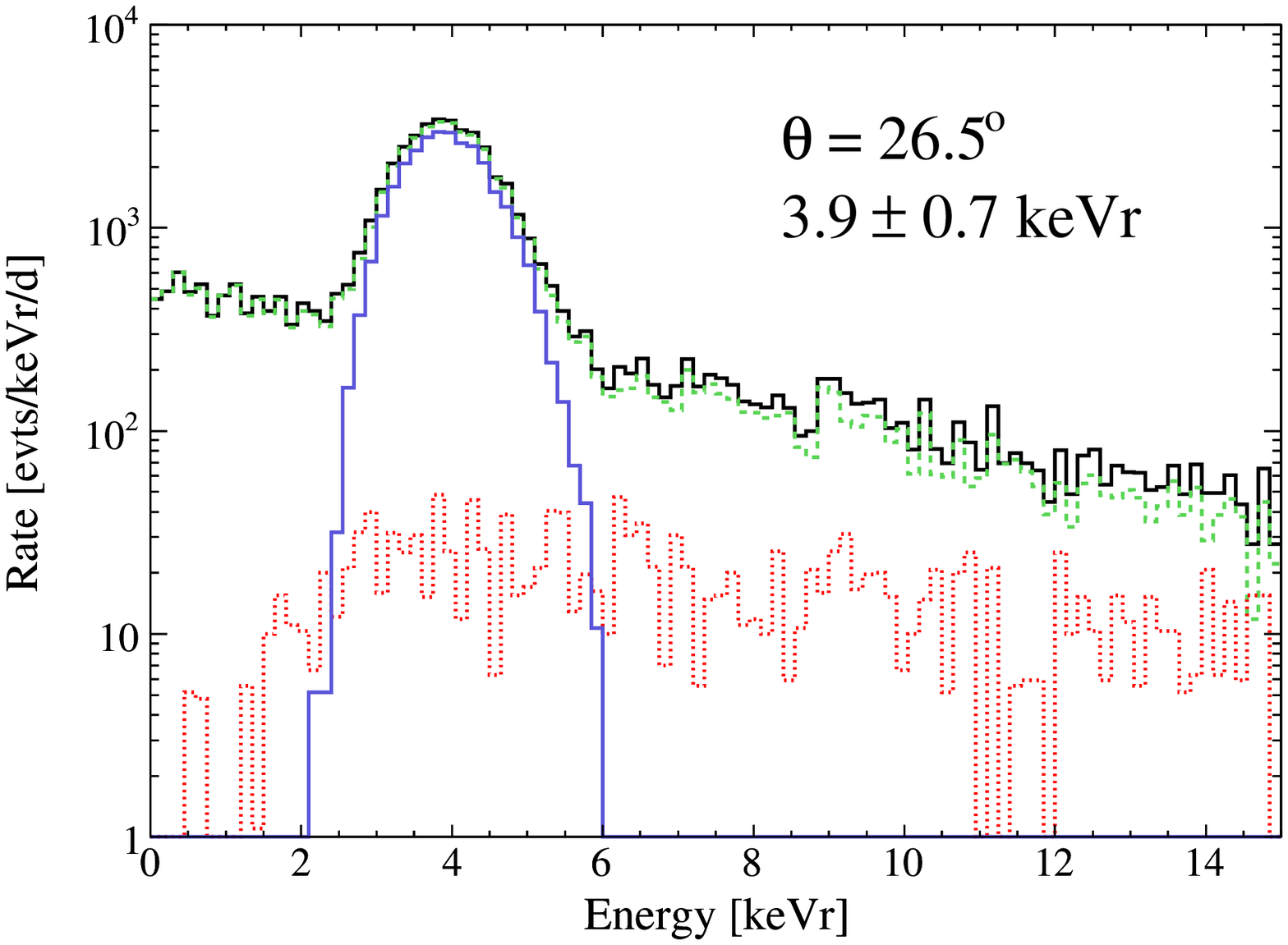} &
		\includegraphics[width=0.9\columnwidth]{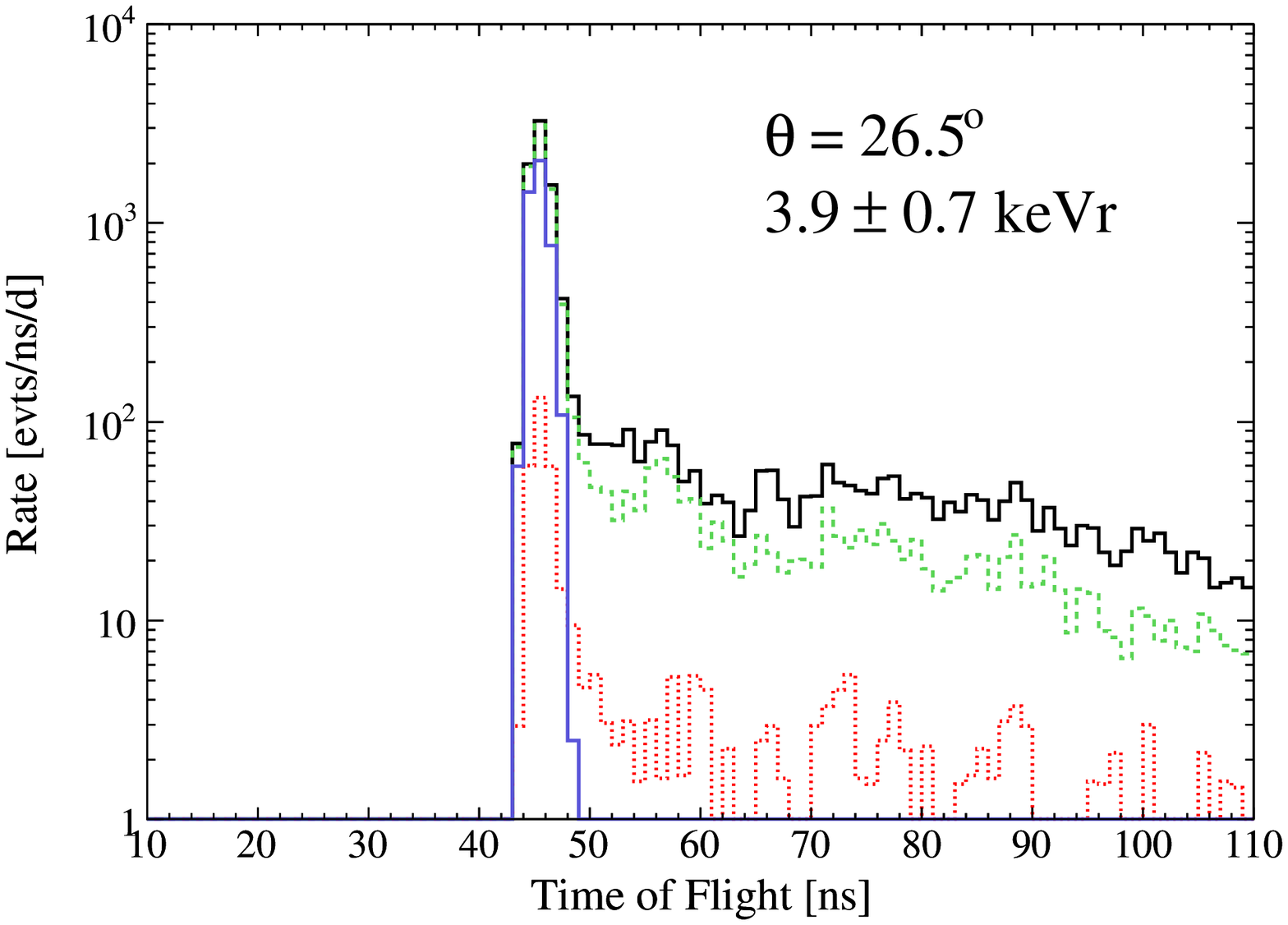} \\
		\includegraphics[width=0.9\columnwidth]{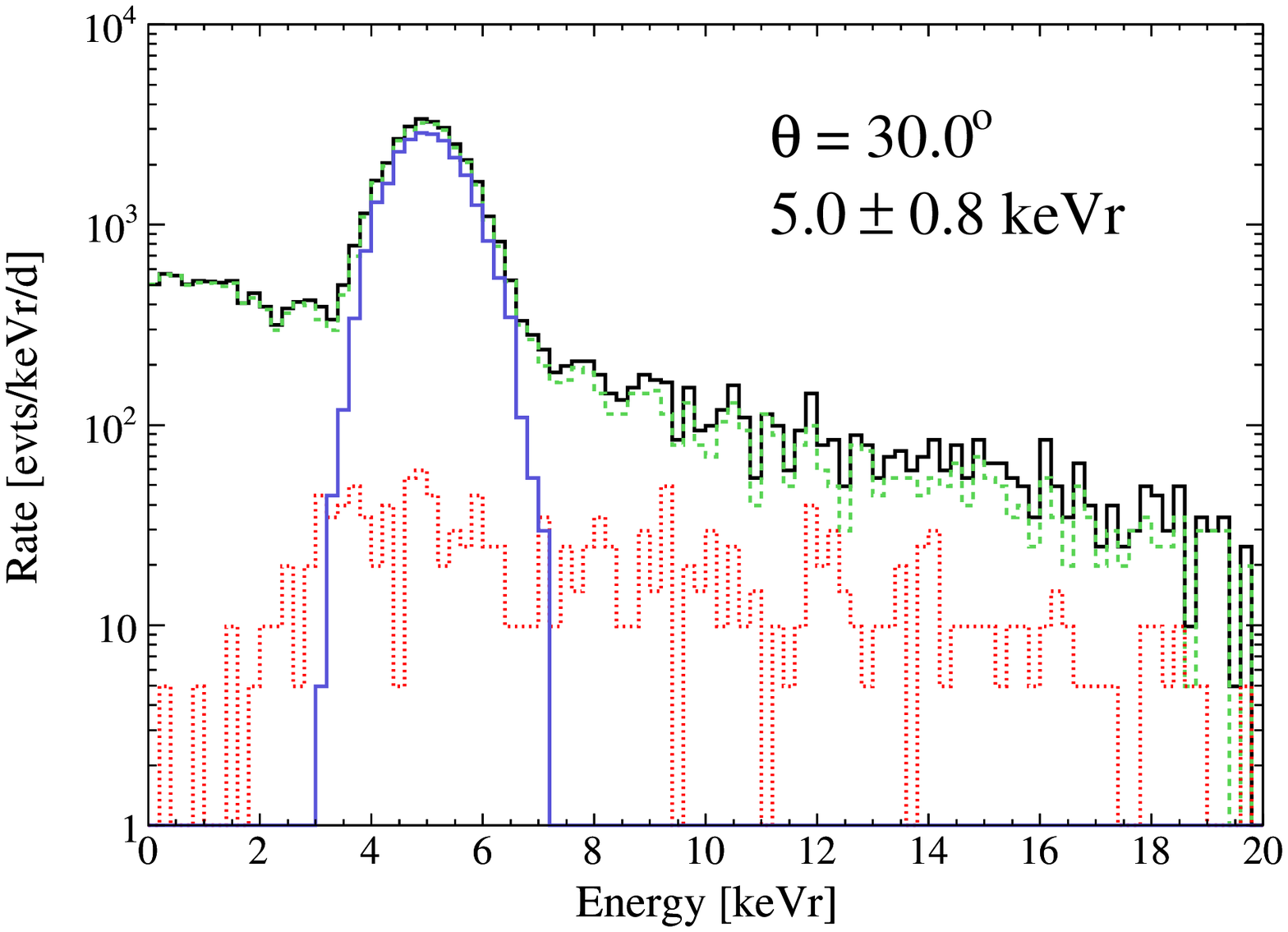} &
		\includegraphics[width=0.9\columnwidth]{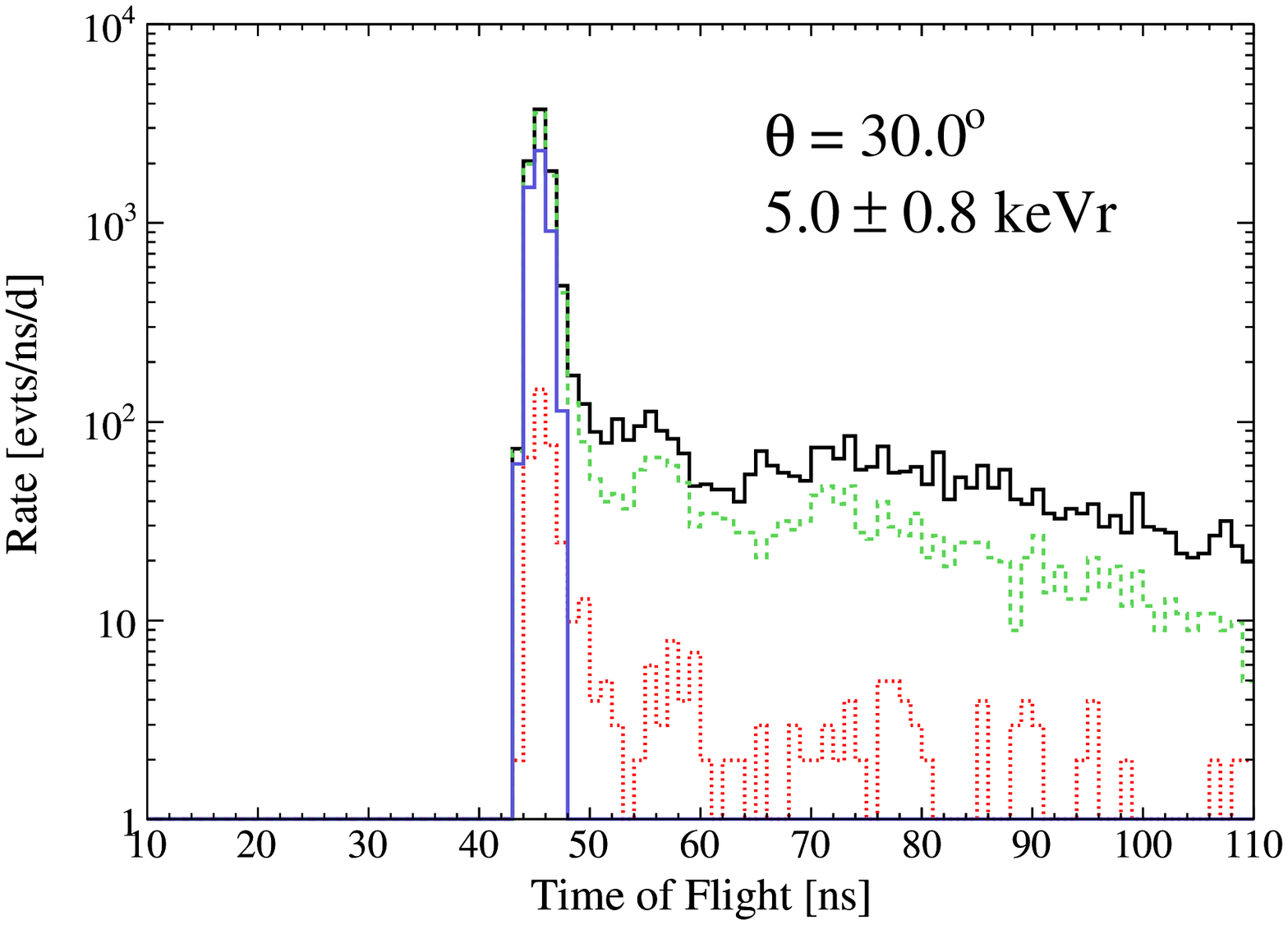} \\
		\includegraphics[width=0.9\columnwidth]{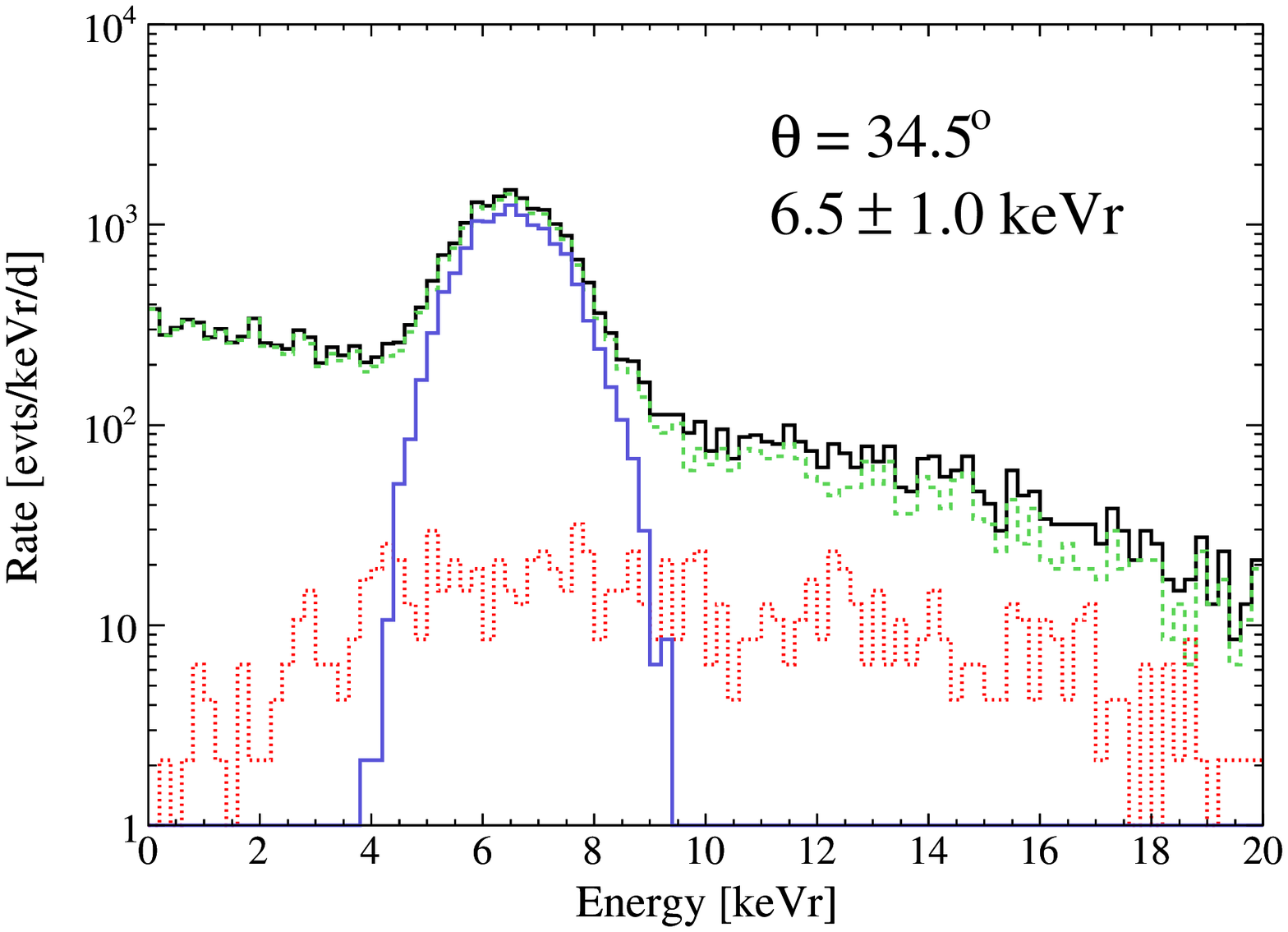} &
		\includegraphics[width=0.9\columnwidth]{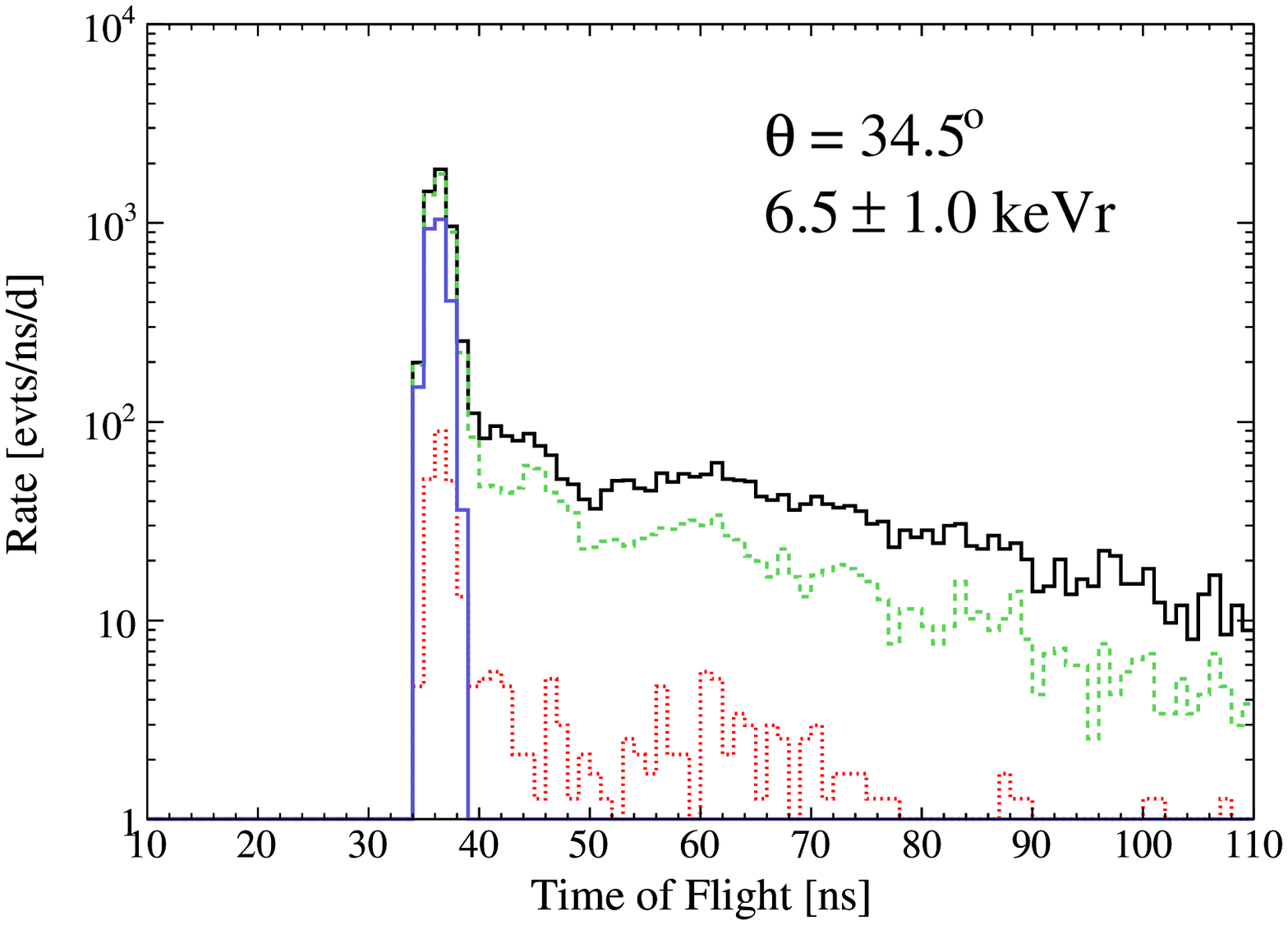} \\
	\end{tabular}
\caption{(left) Simulated LXe nuclear recoil energy spectrum of neutrons interacting in the active LXe volume
and in the EJ301 neutron detector, not including accidental coincidences, for the $23^\circ$, $26.5^\circ$,
$30^\circ$, and $34.5^\circ$ scattering angles, and for the TOF window used in the analysis; and TOF spectrum
(right). The solid black histogram is the total spectrum while the green dashed histogram is the spectrum of
neutrons that elastically scatter once in the LXe active volume and maybe elsewhere. The red dotted histogram
is the spectrum of neutrons that elastically scatter multiple times in the active volume. The blue histogram
is the spectrum of neutrons that interact only via a single elastic scatter in the active LXe volume.}
\label{fig:mc1}
\end{center}
\end{figure*}

\begin{figure*}[htbp]
\begin{center}
	\begin{tabular}{c c}
		\includegraphics[width=0.9\columnwidth]{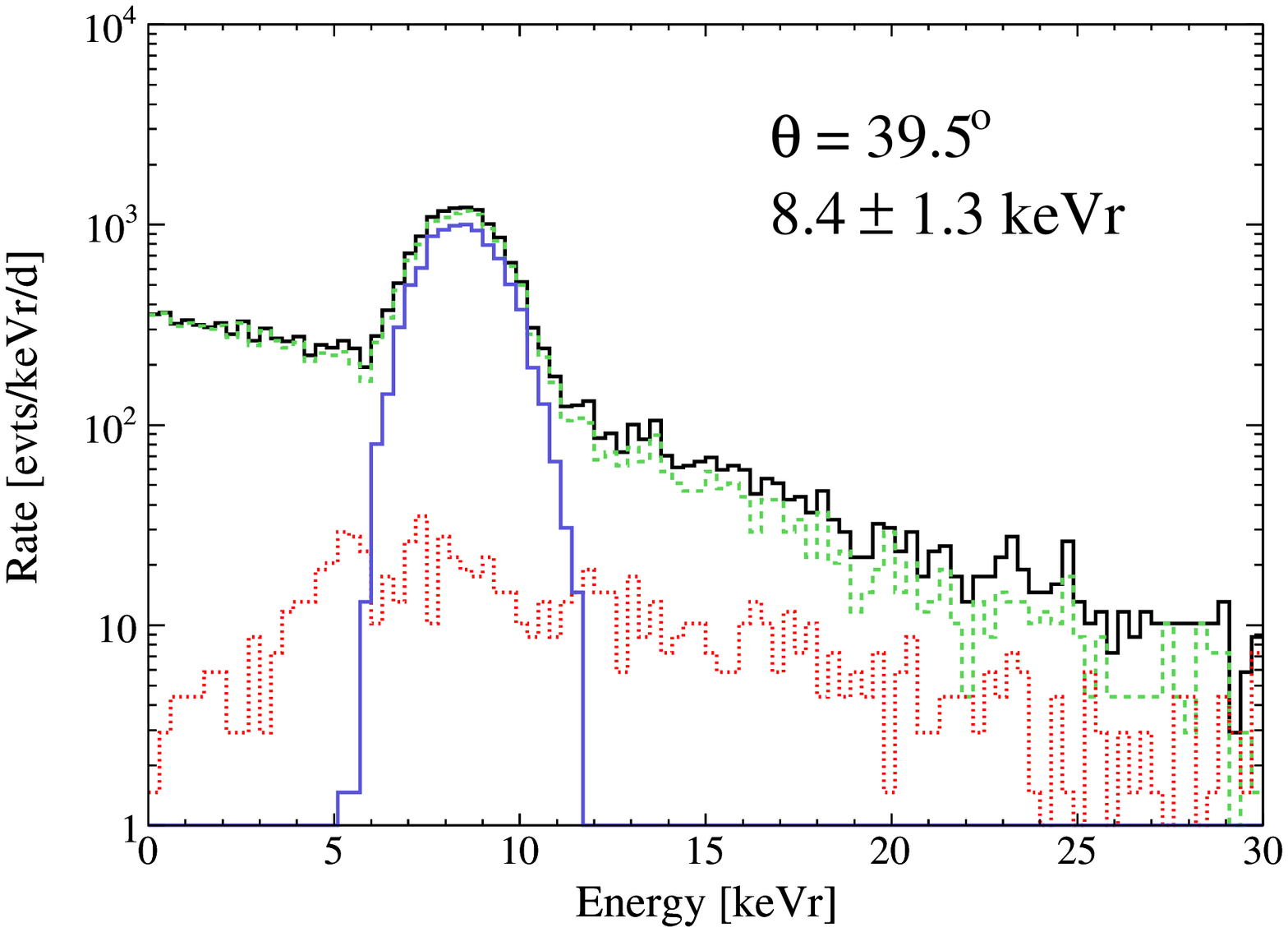} &
		\includegraphics[width=0.9\columnwidth]{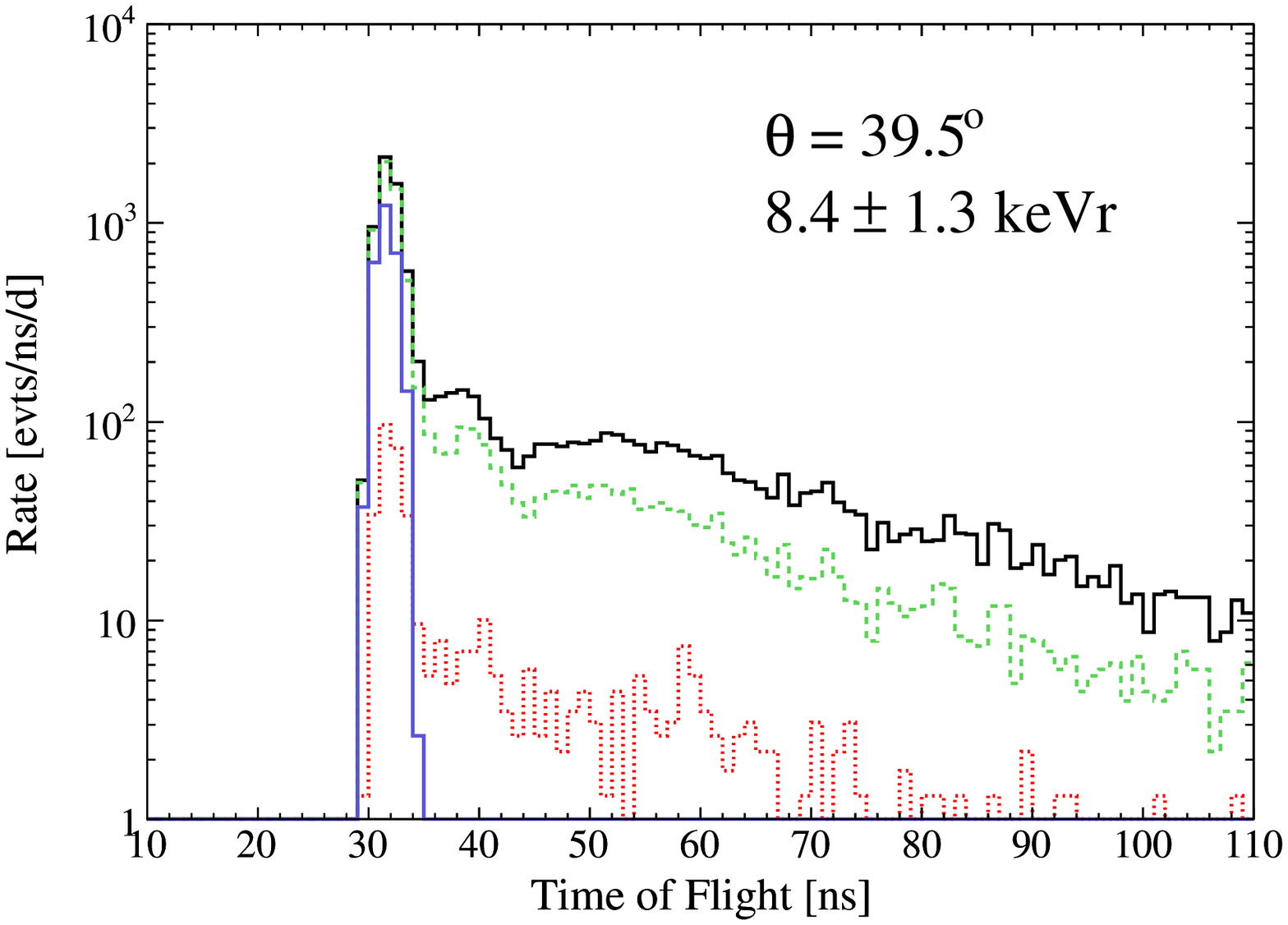} \\
		\includegraphics[width=0.9\columnwidth]{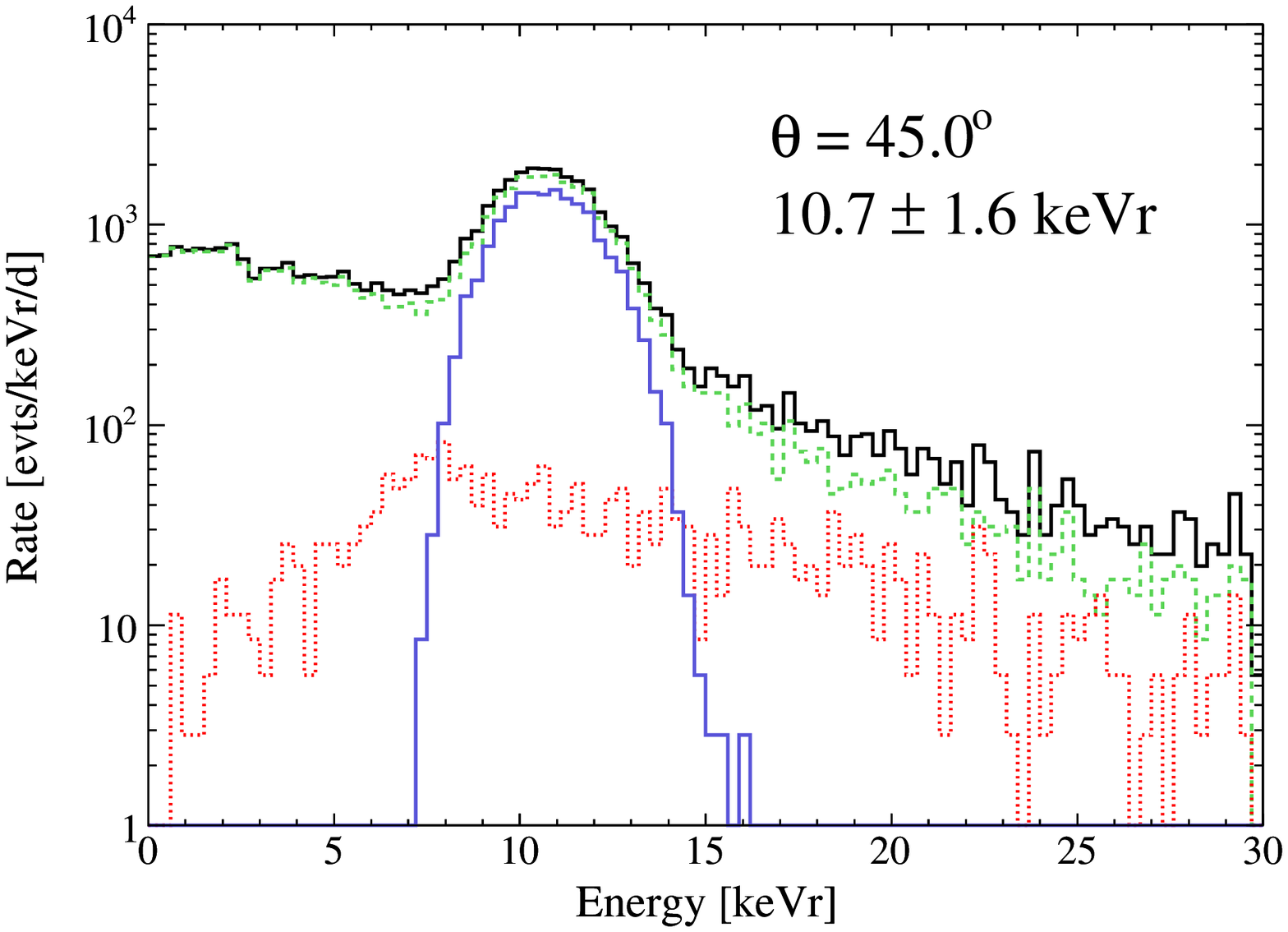} &
		\includegraphics[width=0.9\columnwidth]{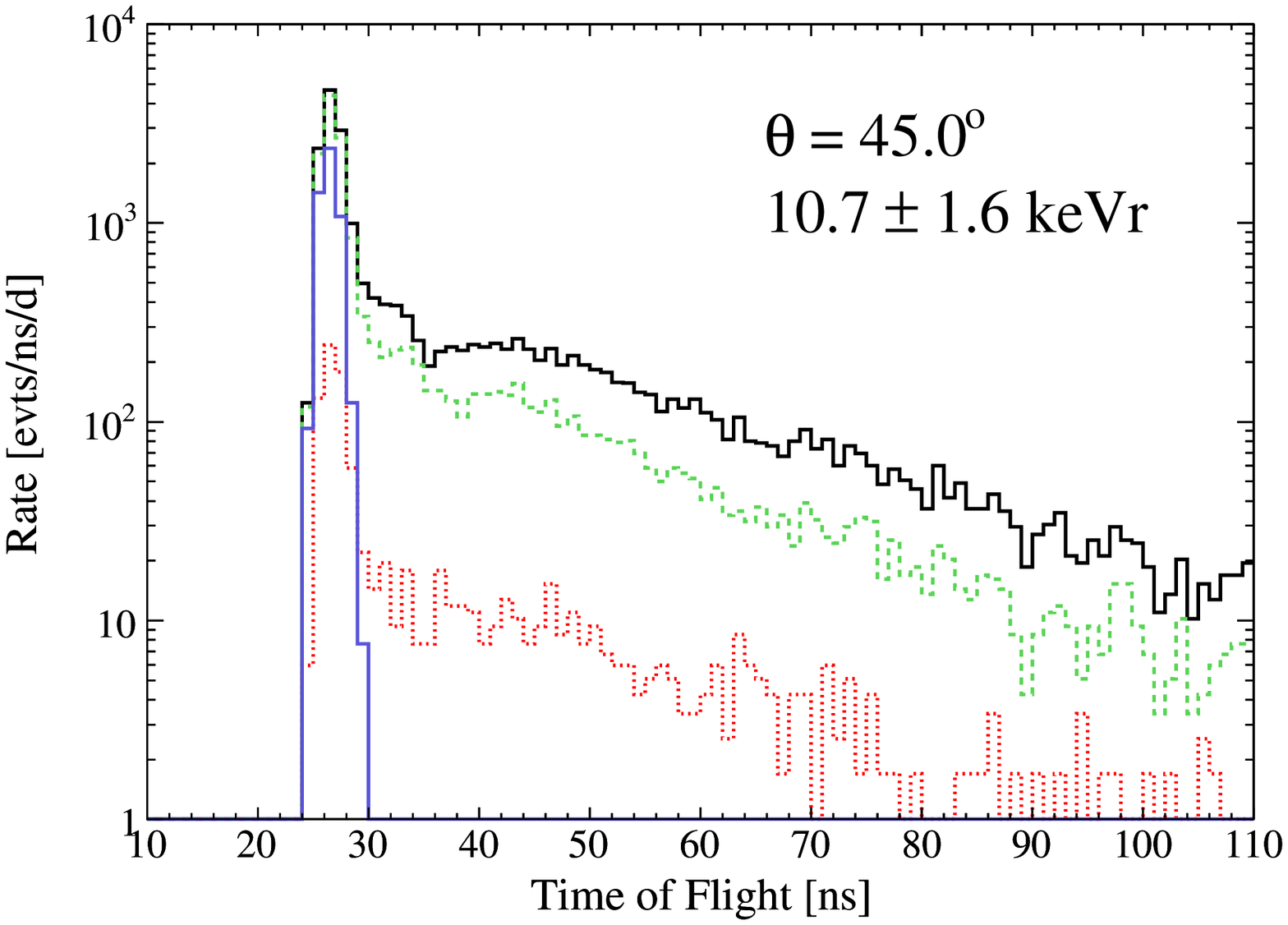} \\
		\includegraphics[width=0.9\columnwidth]{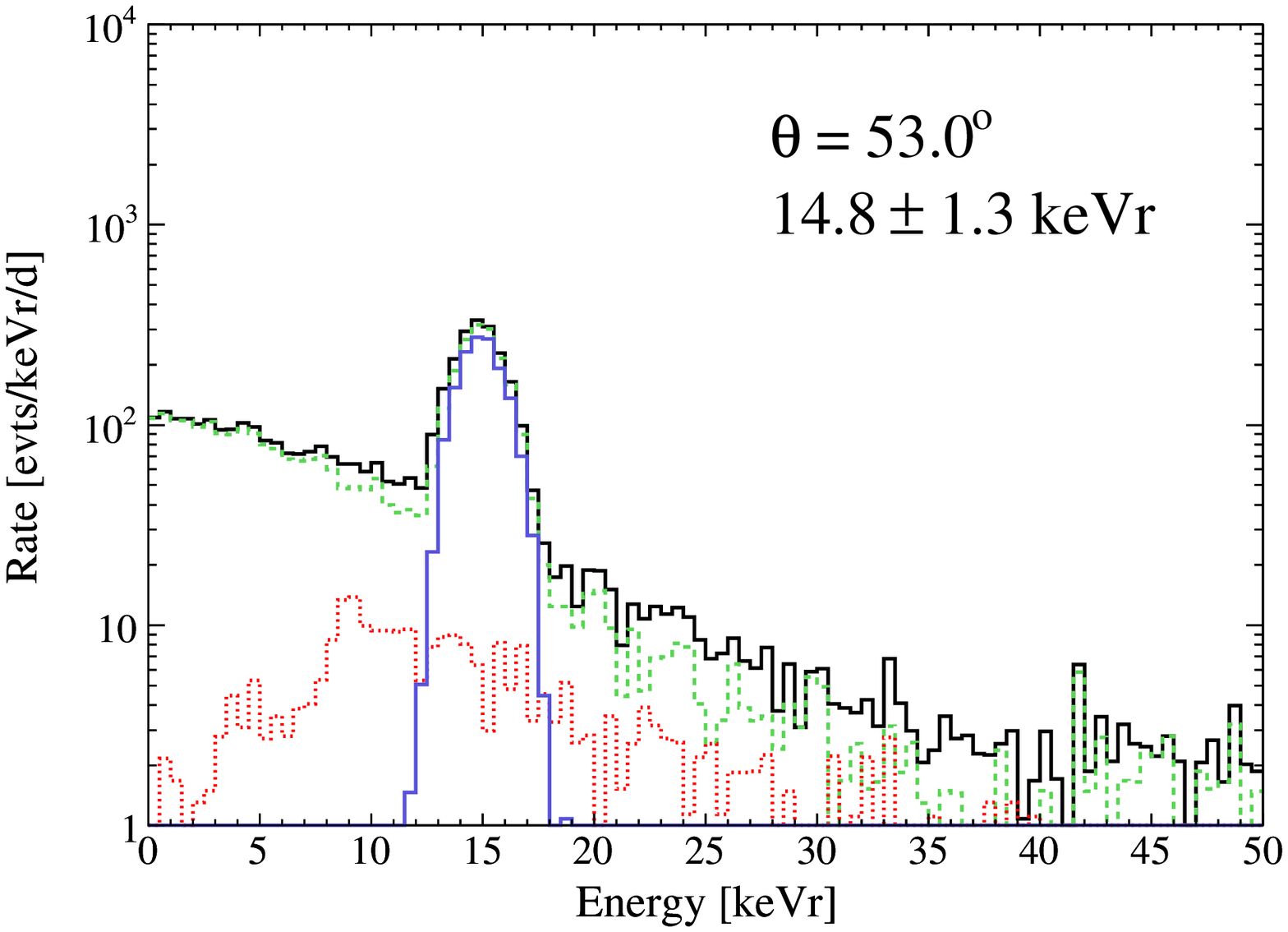} &
		\includegraphics[width=0.9\columnwidth]{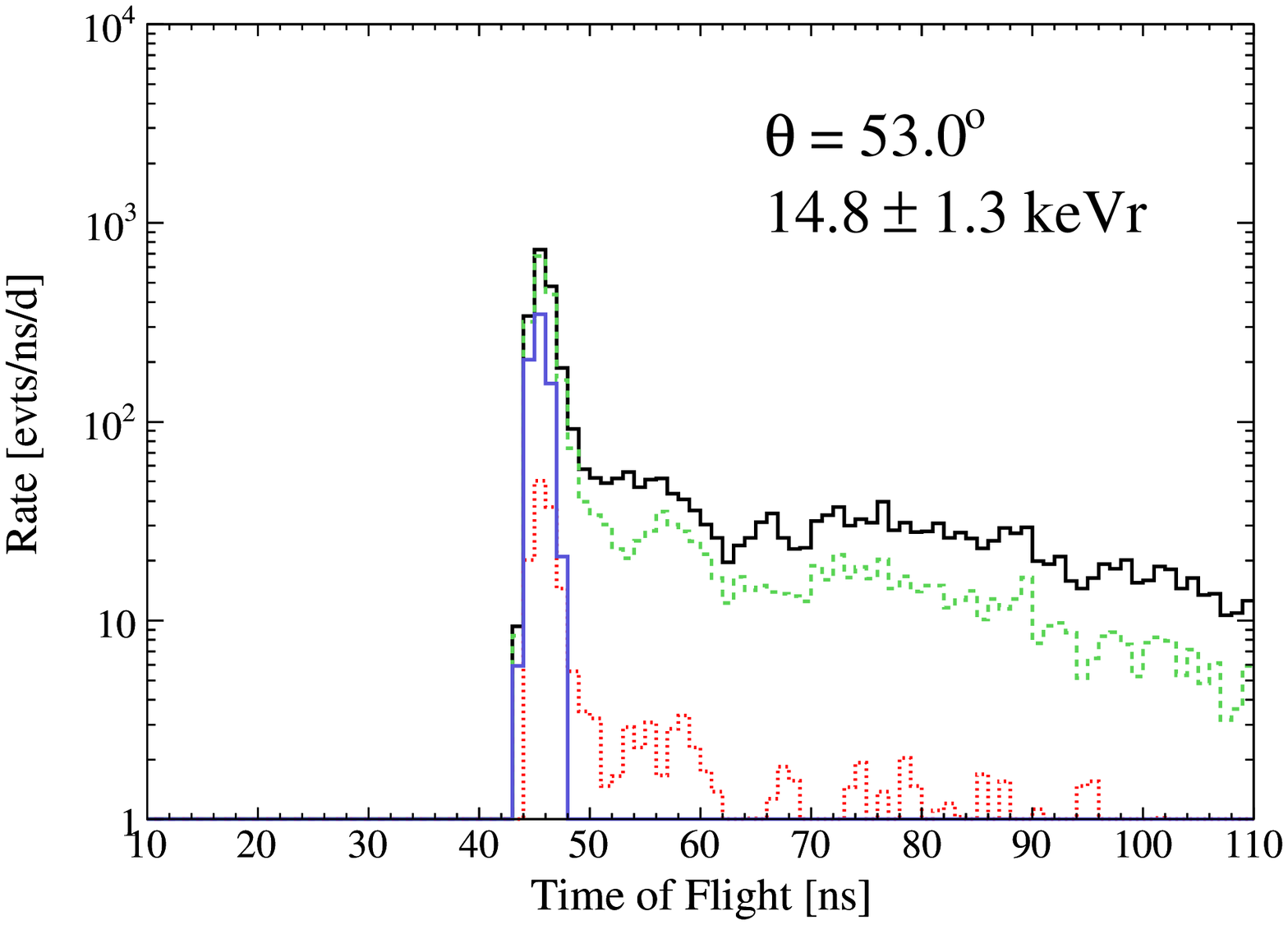} \\
		\includegraphics[width=0.9\columnwidth]{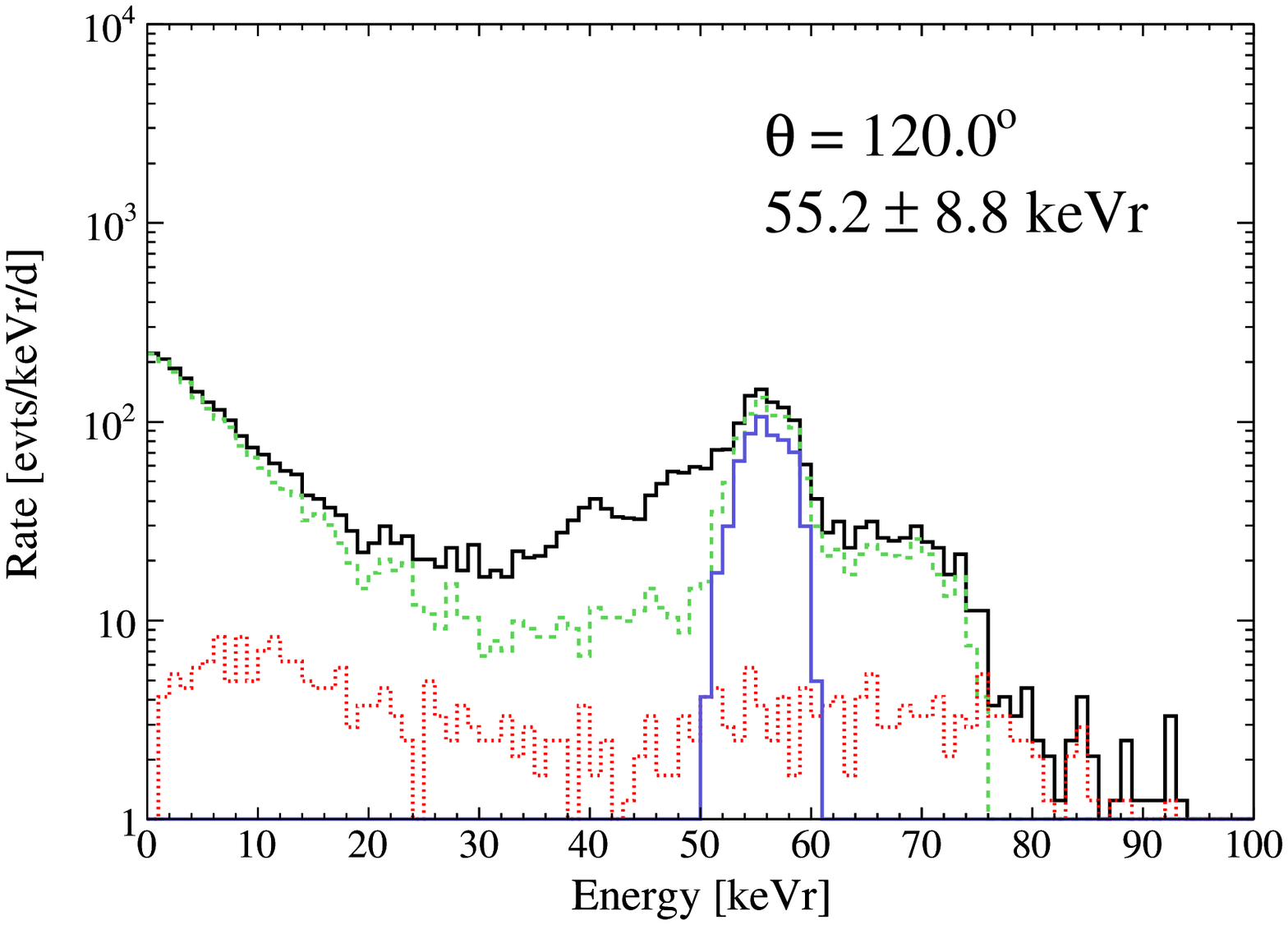} &
		\includegraphics[width=0.9\columnwidth]{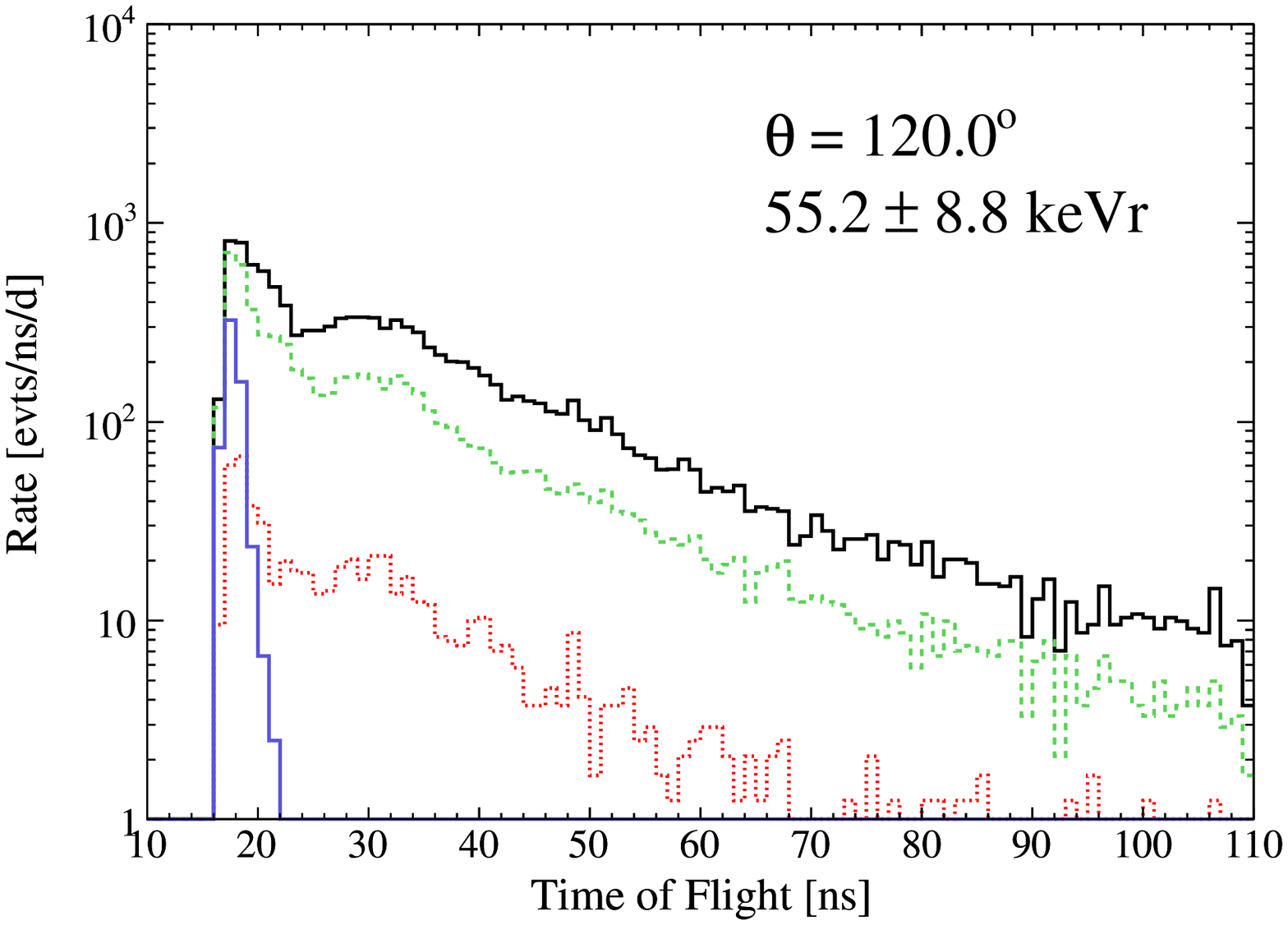} \\
	\end{tabular}
\caption{(left) Simulated LXe nuclear recoil energy spectrum of neutrons interacting in the active LXe volume
and in the EJ301 neutron detector, not including accidental coincidences, for the $39.5^\circ$, $45^\circ$,
$53^\circ$, and $120^\circ$ scattering angles, and for the TOF window used in the analysis; and TOF spectrum
(right). The solid black histogram is the total spectrum while the green dashed histogram is the spectrum of
neutrons that elastically scatter once in the LXe active volume and maybe elsewhere. The red dotted histogram
is the spectrum of neutrons that elastically scatter multiple times in the active volume. The blue histogram
is the spectrum of neutrons that interact only via a single elastic scatter in the active LXe volume.}
\label{fig:mc2}
\end{center}
\end{figure*}

Neutrons that interact in the active LXe volume can deposit energy via elastic or inelastic scattering, once
or multiple times, and may additionally scatter in materials outside of the active volume. The contributions
from all these classes of events has been inferred from the Monte Carlo simulation.
Figs.~\ref{fig:mc1} and~\ref{fig:mc2}
show the simulated recoil energy and TOF distributions for all measured scattering
angles. The energy spectrum of elastic recoils consists of a peak, roughly centered at the recoil energy
corresponding to the angle at which the EJ301 neutron detectors have been placed, and an approximately
exponentially distributed background. The peak is due to neutrons that elastically scatter once in the LXe
detector and interact nowhere else (pure single elastic scatters), while the exponential background is due
to neutrons that additionally scatter in other materials surrounding the active volume, that is, their energy
deposit in the LXe is essentially that of a recoil with a random scattering angle. For the smaller angles the
nuclear recoil energy spectrum is clearly dominated by pure single elastic recoils with 61\% at $23^\circ$ (3
keV) but the proportion gradually decreases reaching 49\% at 8.4~keV and 17.8\% at 55.2~keV. The multiple
scatter contribution is negligible at all angles, from 3.7\% at $23^\circ$ up to 7.0\% at $120^\circ$, due to
the small dimensions of the active LXe volume compared to the neutron elastic scattering mean free path.
These simulation results demonstrate clearly that the design goal of minimizing the amount of materials in the
vicinity of the active LXe volume has been achieved.

\subsection{Extracting $\mathcal{L}_{\n{eff}}$}
\label{sec:leff}

\begin{figure*}[htbp]
\begin{center}
	\begin{tabular}{c c}
		\includegraphics[width=0.9\columnwidth]{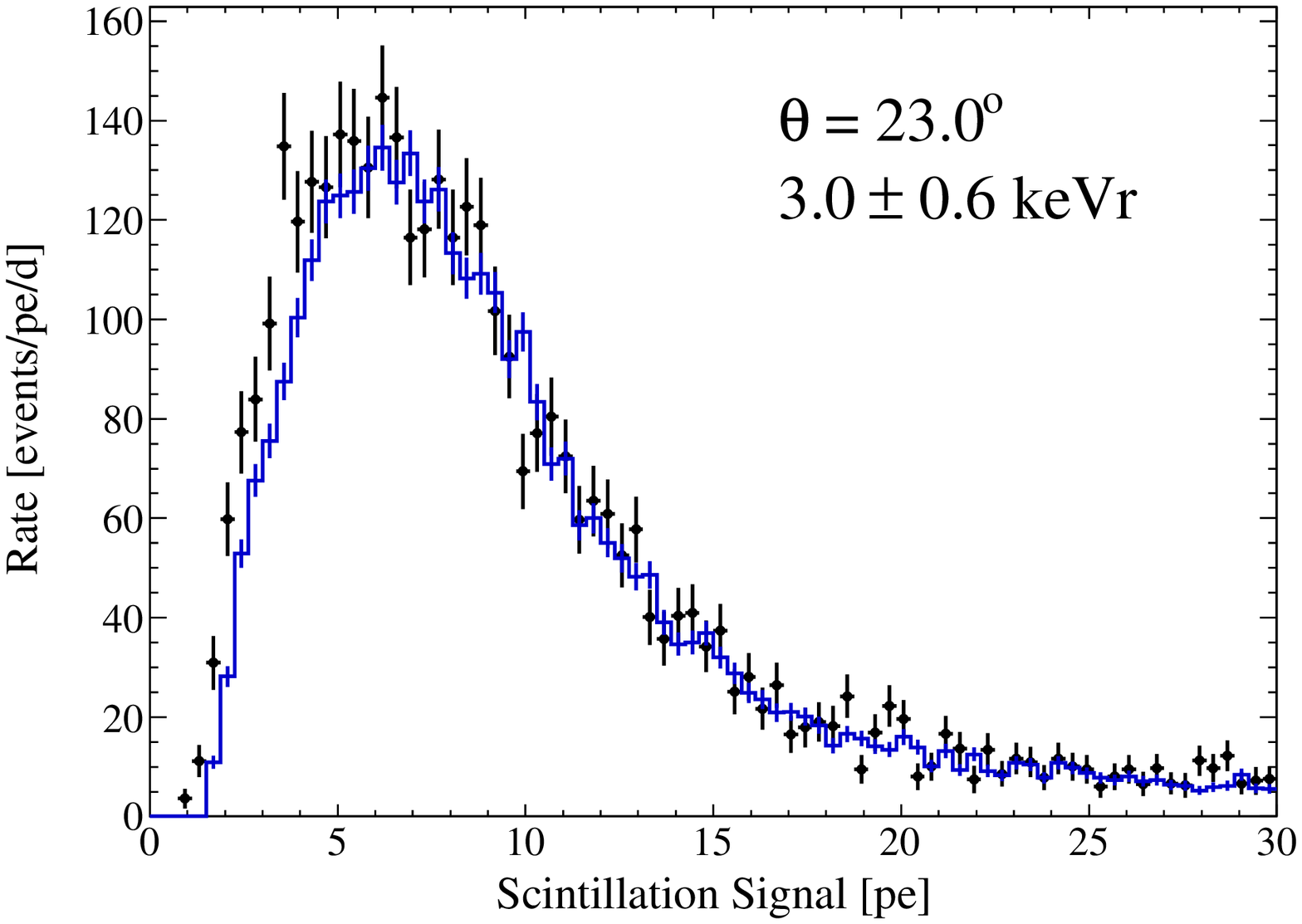} &
		\includegraphics[width=0.9\columnwidth]{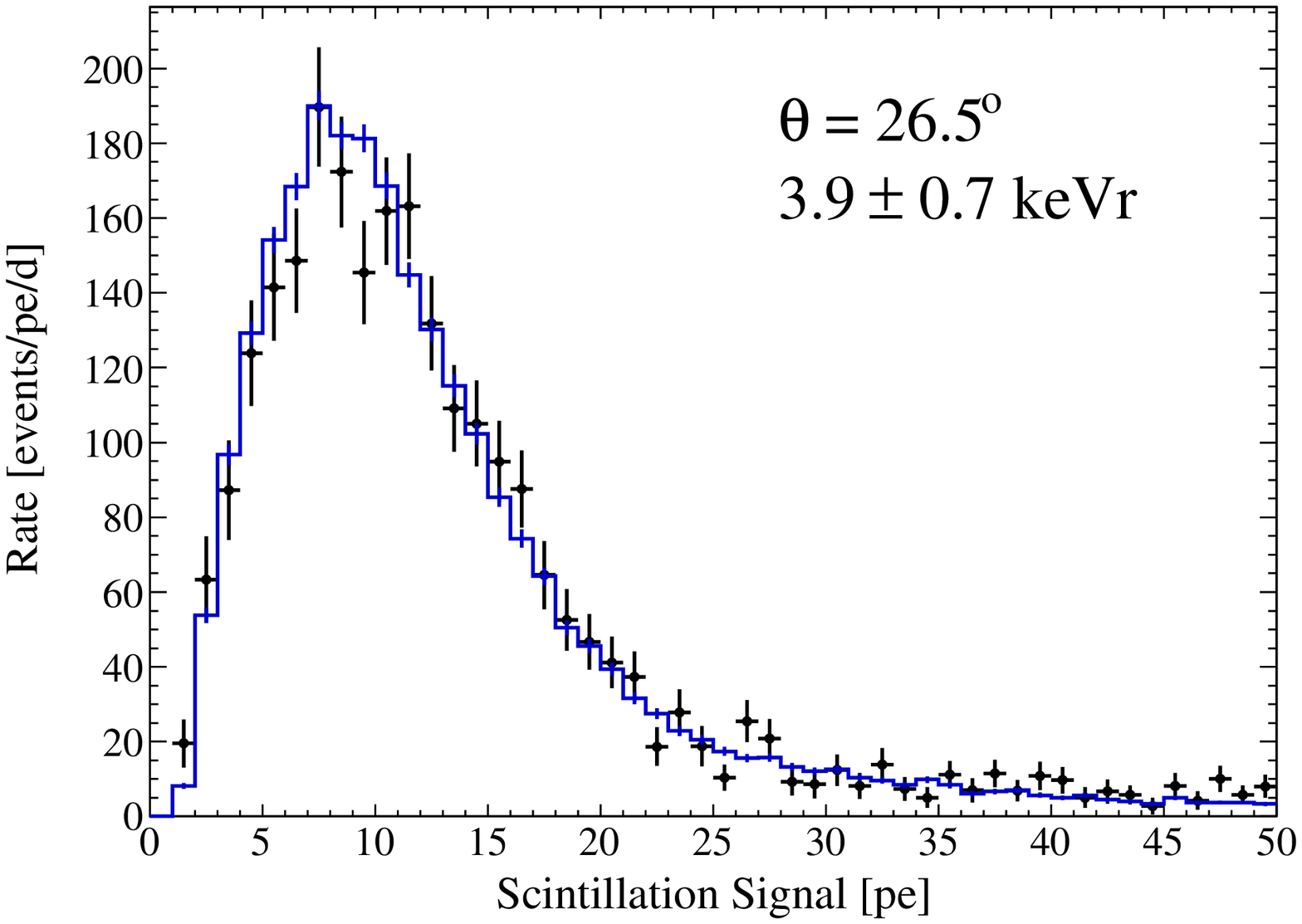} \\
		\includegraphics[width=0.9\columnwidth]{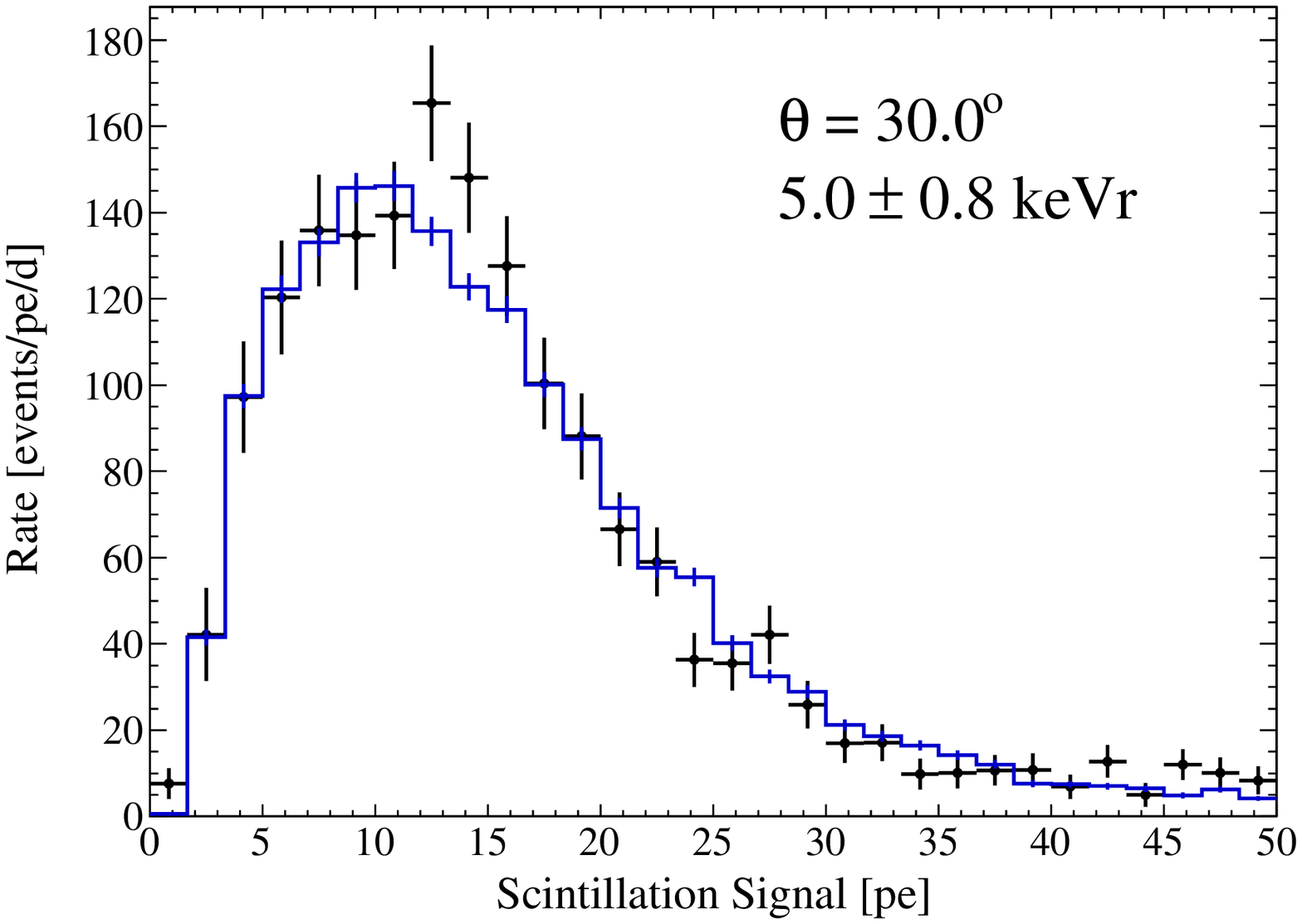} &
		\includegraphics[width=0.9\columnwidth]{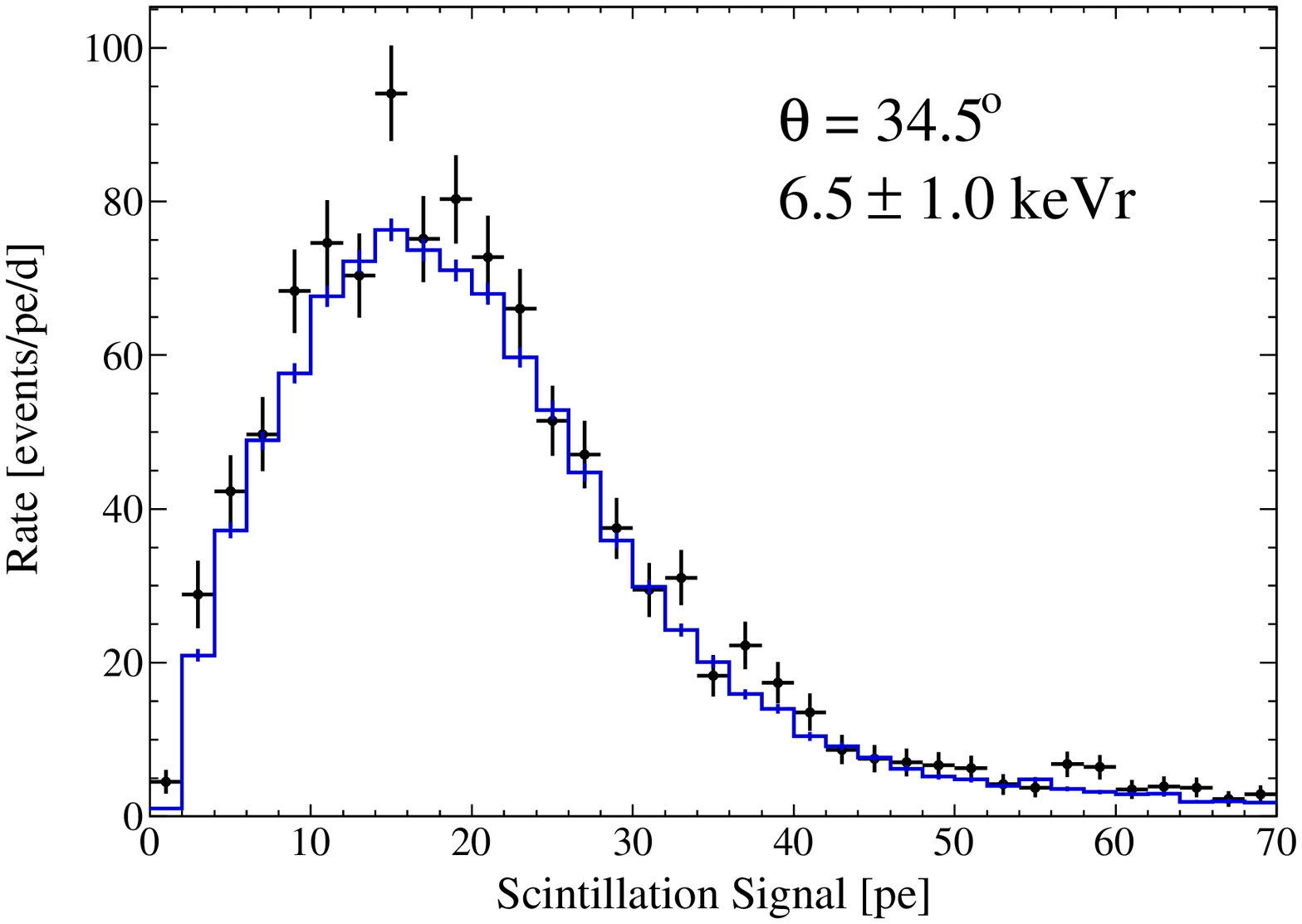} \\
		\includegraphics[width=0.9\columnwidth]{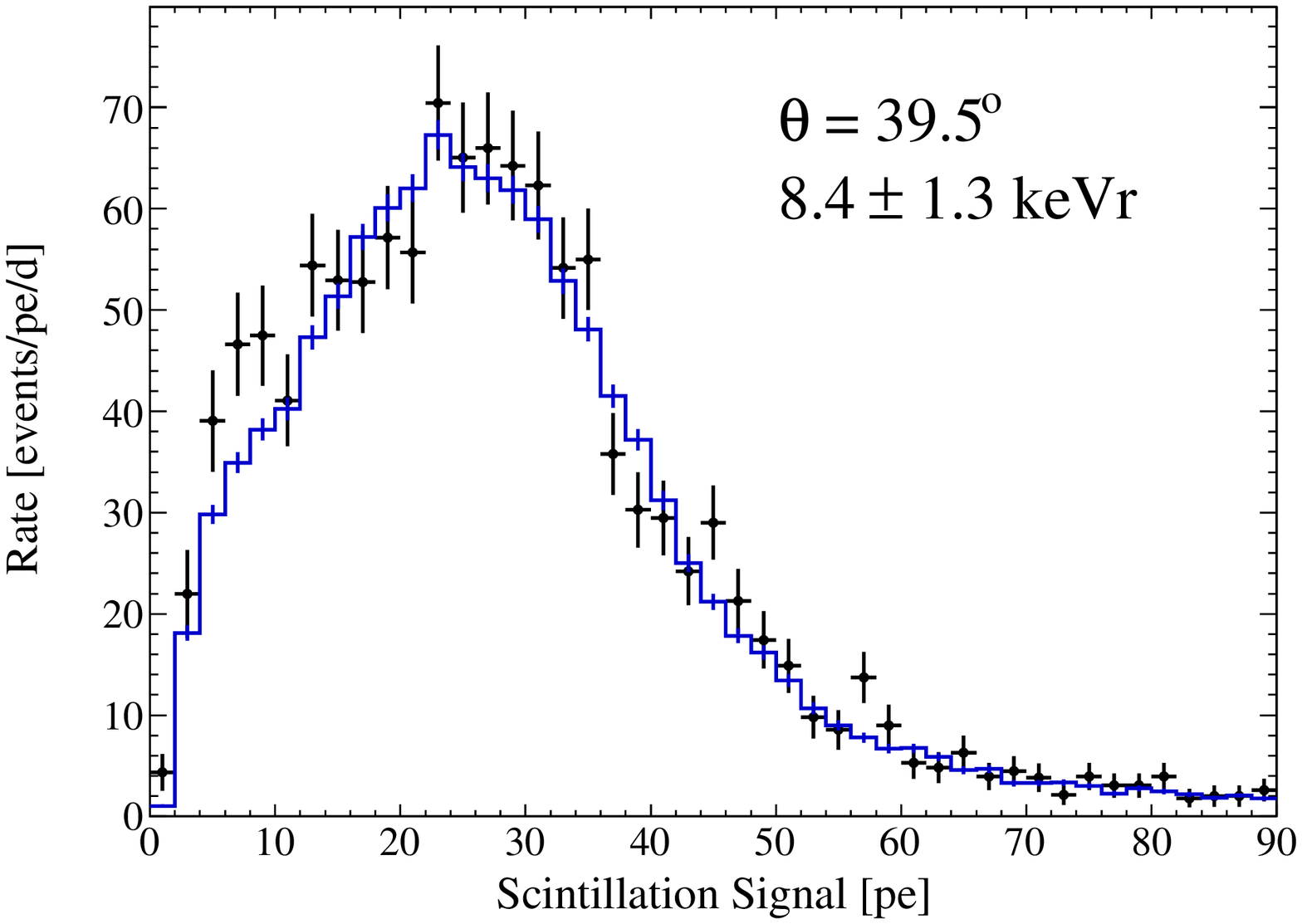} &
		\includegraphics[width=0.9\columnwidth]{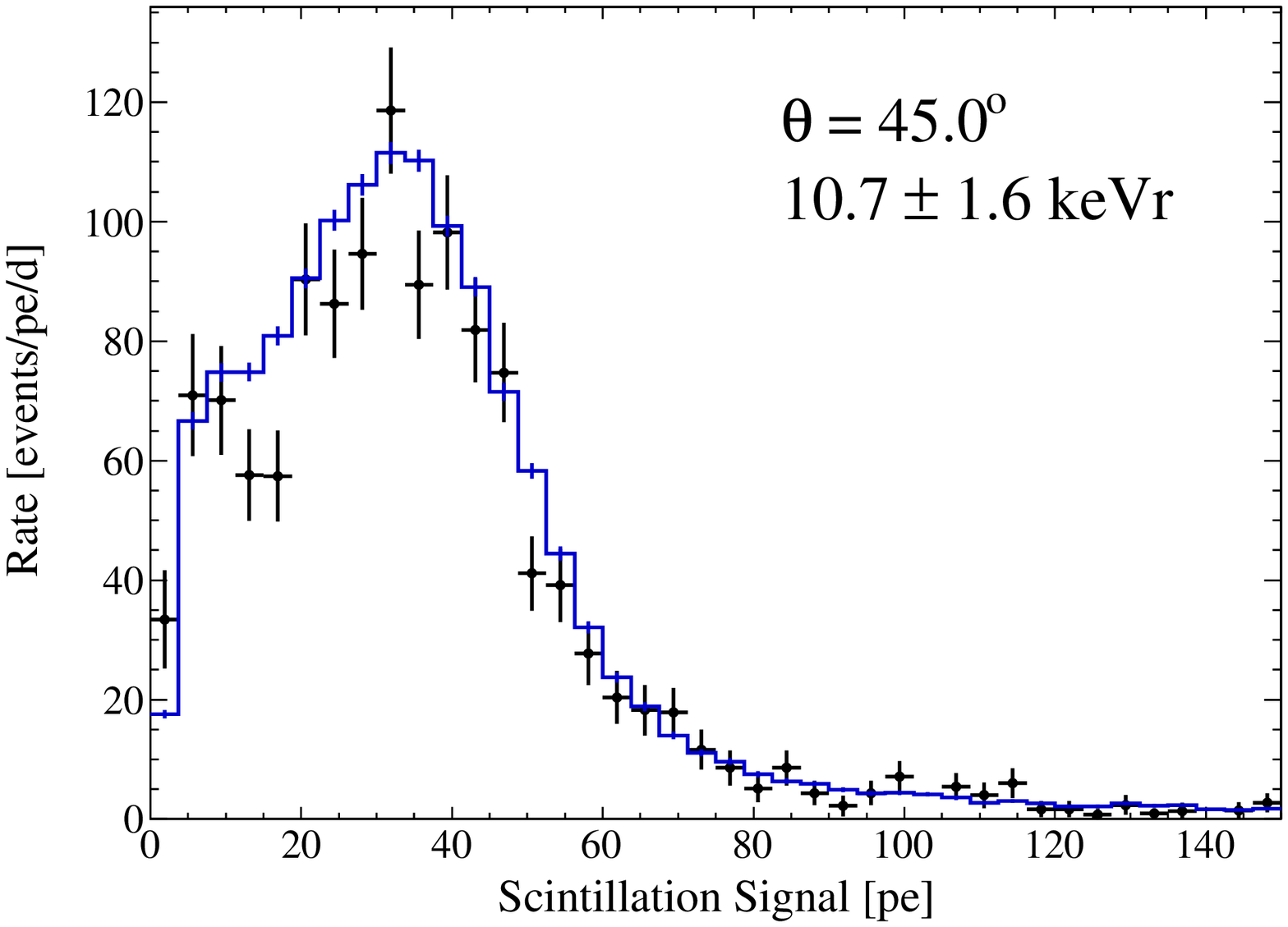} \\
		\includegraphics[width=0.9\columnwidth]{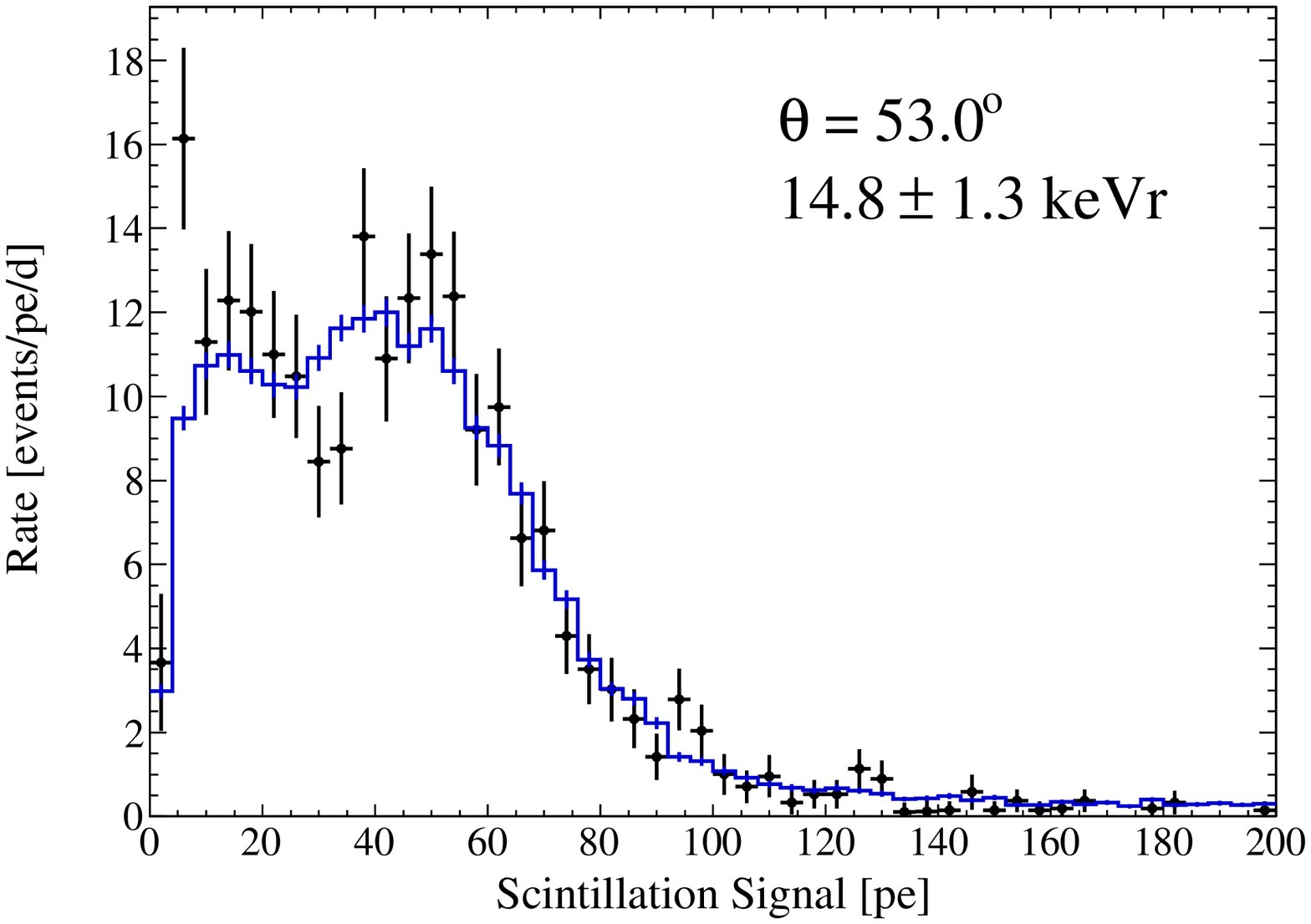} &
		\includegraphics[width=0.9\columnwidth]{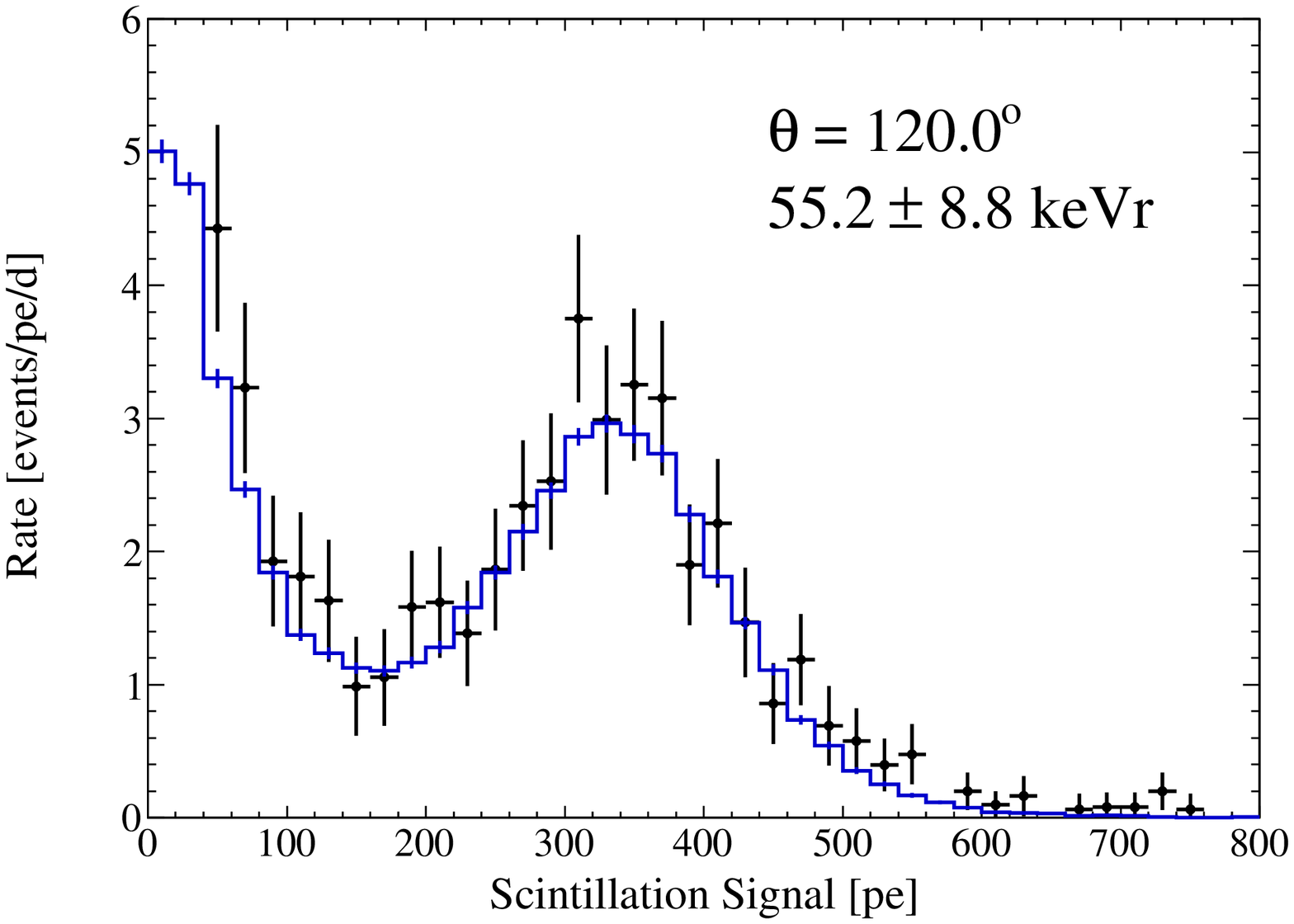} \\
	\end{tabular}
\caption{Fits of Monte Carlo generated recoil spectra (solid lines) to the measured recoil spectra (data points) for all measured scattering angles.}
\label{fig:spectra}
\end{center}
\end{figure*}

% chi^2 analysis
The energy dependence of $\mathcal{L}_{\n{eff}}$ is obtained by minimizing the $\chi^2$
statistic between the measured recoil distribution and the simulated distribution with respect to two free
parameters, $\mathcal{L}_{\n{eff},j} \equiv \mathcal{L}_{\n{eff}}\!\left({E_{\n{nr},j}}\right)$ and $R_j \equiv
R\!\left({E_{\n{nr},j}}\right)$, respectively the scintillation efficiency and the energy resolution at the recoil
energy measured $E_{\n{nr},j}$. These fits are shown in Fig.~\ref{fig:spectra} for all measured scattering angles.
Explicitly, the $\chi^2$ statistic is computed from
\begin{equation}
	\chi^2\!\left({\mathcal{L}_{\n{eff}, j}, R_j}\right) = \sum_{i = 0}^N \frac{\left[{h_i -
	g_i\!\left({\mathcal{L}_{\n{eff}, j}, R_j}\right)}\right]^2}{\sigma_{h,i}^2
	+ \sigma_{g,i}^2\left({\mathcal{L}_{\n{eff}, j}, R_j}\right)}
\end{equation}
where $h_i$ and $g_i$ are the measured and simulated event rates in energy bin $i$, respectively, and $\sigma_{h,i}$
and $\sigma_{g,i}$ the uncertainties in the measured and simulated event rates in bin $i$, respectively.
The bins over which the $\chi^2$ statistic is computed varies depending on the scattering angle so that the
$\chi^2$ does not become dominated by effects in the higher energy tail of the recoil distribution, nor the low
trigger efficiency region. Bins with no counts are removed from the sum (and the number of degrees of
freedom is decreased by one).

The steps involved in transforming the simulated recoil energy distributions of
Figs.~\ref{fig:mc1} and~\ref{fig:mc2}
into the simulated recoil energy distribution $h$, in photoelectrons, are detailed below.

The recoil energy spectrum obtained from the simulation is first multiplied by the $\mathcal{L}_{\n{eff}}$
value under test to convert it to a spectrum with energies in keV (electron-equivalent)
and convolved with a Gaussian energy
resolution with standard deviation $R \sqrt{E}$, where $R$ is the resolution parameter under test.  Next, the
recoil energy spectrum is multiplied by the measured light yield $L_y$ to obtain a spectrum in photoelectrons.
The number of photoelectrons $N_{\n{pe}}$ is allowed to fluctuate according to a Poisson distribution. The
effect of the PMT gain fluctuations is incorporated by convolving the recoil energy spectrum in photoelectrons
with a Gaussian single photoelectron resolution with $0.6 \sqrt{N_{\n{pe}}}$ standard deviation, where $0.6$
is the measured mean PMT single photoelectron resolution. The measured trigger efficiency function
discussed in section~\ref{sec:setup} is then applied to the recoil energy spectrum. As mentioned earlier, the
resulting recoil energy spectrum is divided by the simulation livetime, computed from the neutron generator
yield at the operating conditions. Since $R$ is left as a free parameter during the $\chi^2$
minimization, any additional contribution to the resolution not accounted for will be absorbed in that
parameter.

The last step involves multiplying the simulated recoil energy spectrum by an overall, energy independent
efficiency $\epsilon$, taken as the same for all scattering angle measurements, mostly due to the EJ301 energy
threshold cut and to the uncertainty in the neutron generator yield. This efficiency is computed during the
$\chi^2$ minimization as an additional parameter for the measurement at the recoil energies of $8.4$, $10.7$
and $14.8$ keV. The best fit value from these three measurements is taken as the efficiency for all
measurements, while its uncertainty is taken as the maximum deviation in the measurements. The value obtained
is $\epsilon = 0.41^{+0.04}_{-0.05}$. The 10\% relative uncertainty on the efficiency reflects the fact that
uncertainties in the neutron yield from the generator are at this level.

\section{Results}

\begin{figure}[!htb]
\begin{center}
	\includegraphics[width=1.0\columnwidth]{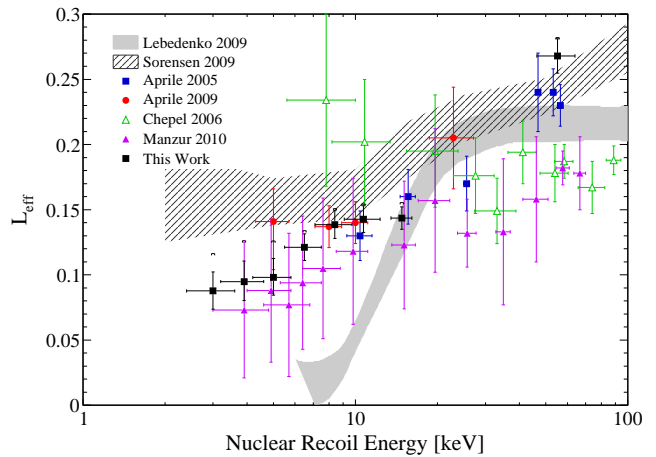} \caption{Measured
	$\mathcal{L}_{\n{eff}}$ values as function of nuclear recoil energy, together with measurements from other
	groups~\cite{Aprile:2005mt,Chepel:2006yv,Aprile:2008rc,Manzur:2009hp,Sorensen:2008ec,Lebedenko:2008gb}. A
	possible additional systematic uncertainty from the trigger efficiency roll-off alone, as described in the
	text, is indicated by the additional error markers.}\label{fig:results}
\end{center}
\end{figure}

% leff results
The $\mathcal{L}_{\n{eff}}$ values obtained for all scattering angles measured are listed in Table
\ref{tab:results}. Fig.~\ref{fig:results} shows the results along with those of prior measurements at low
energies~\cite{Aprile:2005mt,Chepel:2006yv,Aprile:2008rc,Manzur:2009hp,Sorensen:2008ec,Lebedenko:2008gb}.

\begin{table}[htbp]
	\caption{Values of $\mathcal{L}_{\n{eff}}$ obtained at the 8~angles used in this study,
                 together with their errors as discussed in the text.}\label{tab:results}
	\begin{tabular*}{\columnwidth}{@{\extracolsep{\fill}} l c c}
		\hline \hline
		$\theta$ & $E_{\n{nr}}$ (keV) & $\mathcal{L}_{\n{eff}}$ \\
		\hline
		23$^\circ$ &   $3.0 \pm 0.6$ & $0.088^{+0.014}_{-0.015}$ \\
		26.5$^\circ$ & $3.9 \pm 0.7$ & $0.095^{+0.015}_{-0.016}$ \\
		30$^\circ$ &   $5.0 \pm 0.8$ & $0.098^{+0.014}_{-0.015}$ \\
		34.5$^\circ$ & $6.5 \pm 1.0$ & $0.121 \pm 0.010$ \\
		39.5$^\circ$ & $8.4 \pm 1.3$ & $0.139 \pm 0.011$ \\
		45$^\circ$ &  $10.7 \pm 1.6$ & $0.143 \pm 0.010$ \\
		53$^\circ$ &  $14.8 \pm 1.3$ & $0.144 \pm 0.009$ \\
		120$^\circ$ & $55.2 \pm 8.8$ & $0.268 \pm 0.013$ \\
		\hline \hline
	\end{tabular*}
\end{table}

The total uncertainty on $\mathcal{L}_{\n{eff}}$ is given by a combination of statistical and systematic
factors with the statistical uncertainty taken from the fit to the data. The systematic uncertainties include
contributions from the spread in nuclear recoil energies, $\sigma_{E_{\n{nr}}}$, and uncertainties associated
with the $^{57}\n{Co}$ light yield, $\sigma_{L_y}$, the efficiency of the liquid scintillator threshold cut,
$\sigma_{\epsilon}$, and the positions of the neutron generator, $\sigma_{r_g}$, and of the EJ301 detectors,
$\sigma_{r_s}$. Explicitly
\begin{multline}
	\sigma^2_{\mathcal{L}_{\n{eff}}} = \sigma^2_{\mathcal{L}_{\n{eff}}, \n{fit}}
	+ \bigl({\tfrac{\partial \mathcal{L}_{\n{eff}}}{\partial L_y}}\bigr)^2 \sigma^2_{L_y}
	+ \bigl({\tfrac{\partial \mathcal{L}_{\n{eff}}}{\partial E_{\n{nr}}}}\bigr)^2 \sigma^2_{E_{\n{nr}}} \\
	+ \bigl({\tfrac{\Delta \mathcal{L}_{\n{eff}}}{\Delta \epsilon}}\bigr)^2 \sigma^2_{\epsilon}
	+ \bigl({\tfrac{\Delta \mathcal{L}_{\n{eff}}}{\Delta r_g}}\bigr)^2 \sigma^2_{r_g}
	+ \bigl({\tfrac{\Delta \mathcal{L}_{\n{eff}}}{\Delta r_s}}\bigr)^2 \sigma^2_{r_s} \n{.}
\end{multline}
The change in $\mathcal{L}_{\n{eff}}$ with nuclear recoil energy, $\partial \mathcal{L}_{\n{eff}}/\partial
E_{\n{nr}}$, is computed in closed form from a logarithmic fit to the measured $\mathcal{L}_{\n{eff}}$ values
versus nuclear recoil energy. The change in the inferred $\mathcal{L}_{\n{eff}}$ values due to the
uncertainties in the efficiency, and the positions of the neutron generator and of the EJ301 detectors,
$\Delta \mathcal{L}_{\n{eff}}/\Delta \epsilon$, $\Delta \mathcal{L}_{\n{eff}}/\Delta r_g$, and $\Delta
\mathcal{L}_{\n{eff}}/\Delta r_s$, respectively, were all calculated through a discrete approximation by
performing additional Geant4 simulations where each parameter is varied by small amounts.

The largest contribution to the total uncertainty in $\mathcal{L}_{\n{eff}}$ comes from the spread in recoil
energy. At energies below 6.5~keV, this is followed by the uncertainty in the efficiency $\epsilon$.
This is expected as in this
region, $\mathcal{L}_{\n{eff}}$ varies most with nuclear recoil energy, and since the mean of the pure
single elastic recoil peak is in the roll-off of the trigger efficiency curve. For energies of 6.5~keV and
above, the statistical uncertainty from the fit is the next-to leading contribution
to the total uncertainty. This is likely caused by the smaller statistics acquired for
the higher recoil energy datasets.

The systematic uncertainty in $\mathcal{L}_{\n{eff}}$ coming from the uncertainty in the trigger efficiency
roll-off has been investigated by varying the trigger efficiency function. If one assumes the measured
trigger efficiency as the true efficiency, then its statistical uncertainty has a negligeable effect on the
inferred $\mathcal{L}_{\n{eff}}$ values. However, if one assumes that a systematic effect is responsible for
the discrepancy between the mesured and simulated trigger efficiencies and takes the simulated efficiency as
the true efficiency, then the effect on the $\mathcal{L}_{\n{eff}}$ values below 6.5~keV is substantial. The
effect is indicated by the additional error indicators in in Fig.~\ref{fig:results}.

\section{Discussion}

Our results suggest that $\mathcal{L}_{\n{eff}}$ slowly decreases with decreasing energy, from $0.144 \pm
0.009$ at 15~keV to $0.088^{+0.014}_{-0.015}$ at 3~keV. The agreement at 8.4 and 10.7~keV with the points at
8~and 10~keV of Aprile~\textit{et. al}~\cite{Aprile:2008rc} is excellent. Considering that the two
measurements were performed with different LXe detectors, using different incident neutron energies, and at
different neutron fluxes, reinforces the accuracy of the new measurement. Our results below 10 keV are
consistent with those of Manzur~\textit{et al.}~\cite{Manzur:2009hp}, within errors, but the decreasing trend
observed is not as pronounced. However, they are incompatible with the indirect $\mathcal{L}_{\n{eff}}$
measurement of Lebedenko~\textit{et. al}~\cite{Lebedenko:2008gb} or with considerably smaller
$\mathcal{L}_{\n{eff}}$ values such as those suggested in~\cite{Collar:2010ht}.

% \todo{The observed decrease in $\mathcal{L}_{\n{eff}}$ is compatible with the semi-empirical model proposed
% in \cite{Manfred}.}

We have not performed any measurement of $\mathcal{L}_{\n{eff}}$ below 3~keV to avoid the low trigger
efficiency region. If $\mathcal{L}_{\n{eff}}$ were to decrease below 3~keV with the same logarithmic
slope as observed between 3~and 10.7~keV, the single elastic recoil peak for a 2~keV measurement, for example,
would lie at a trigger efficiency of 45\%. At this stage, the fraction of the neutron scattering rate
which produces measurable signals becomes comparable to the relative uncertainty on the neutron generator
yield. Considering that this LXe detector has the highest light detection efficiencies achieved in a LXe detector,
measuring $\mathcal{L}_{\n{eff}}$ in the near future at lower energies is probably impractical
and will be subject to a considerably higher systematic uncertainty from the trigger efficiency roll-off.

Given that most LXe dark matter detectors use both the scintillation and ionization signals to distinguish
nuclear recoils from backgrounds and would thus benefit from an improved energy calibration using both
signals, the next step will be to simultaneously measure the scintillation and ionization yield of nuclear
recoils.

\begin{acknowledgements}
	This work was carried out with support from the National Science Foundation for the XENON100 Dark Matter
	experiment (Award No. PHYS09-04220). We express our gratitude to Andrew Bazarko and Peter
	Wraight from the Schlumberger Princeton Technology Center, for their support with the neutron generator.
\end{acknowledgements}

%\bibliographystyle{apsrev}
%\bibliography{xenonleff}

\end{document}